\documentclass[journal]{IEEEtran}
\usepackage{amsmath,graphicx,cite,epsfig}
\usepackage{multirow}
\usepackage{booktabs}
\usepackage{float}
\usepackage{subfigure}
\usepackage{caption}
\usepackage{threeparttable}
\usepackage{diagbox}
\usepackage{url}
\usepackage{bm}
\usepackage{stfloats}

\ifCLASSINFOpdf

\else

\fi

\begin{document}
%
\title{Binocular Rivalry Oriented Predictive \\Auto-Encoding Network for Blind Stereoscopic Image Quality Measurement}
%
%
%

\author{Jiahua Xu, Wei Zhou,~\IEEEmembership{Student~Member,~IEEE}, Zhibo Chen,~\IEEEmembership{Senior~Member,~IEEE}, \\Suiyi Ling,~\IEEEmembership{Member,~IEEE}, and Patrick Le Callet,~\IEEEmembership{Fellow,~IEEE}
\thanks{This work was supported in part by NSFC under Grant U1908209, 61632001 and the National Key Research and Development Program of China 2018AAA0101400 and Alibaba Corporate. (Jiahua Xu and Wei Zhou contributed equally to this work.) (Corresponding anthor: Zhibo Chen.)}
\thanks{J. Xu, W. Zhou and Z. Chen are with the CAS Key Laboratory of Technology in Geo-Spatial Information Processing and Application System, University of Science and Technology of China, Hefei 230027, China (e-mail: xujiahua@mail.ustc.edu.cn; weichou@mail.ustc.edu.cn; chenzhibo@ustc.edu.cn).}
\thanks{S. Ling and P. Le Callet are with the Équipe Image, Perception et Interaction, Laboratoire des Sciences du Numérique de Nantes, Université de Nantes, 44300 Nantes, France (e-mail: suiyi.ling@univ-nantes.fr; patrick.lecallet@univ-nantes.fr).}
\thanks{}}

%
%

\markboth{IEEE Transactions on Instrumentation and Measurement}%
{Shell \MakeLowercase{\textit{et al.}}: Bare Demo of IEEEtran.cls for IEEE Journals}
%



\maketitle

\begin{abstract}
Stereoscopic image quality measurement (SIQM) has become increasingly important for guiding stereo image processing and commutation systems due to the widespread usage of 3D contents. Compared with conventional methods which are relied on hand-crafted features, deep learning oriented measurements have achieved remarkable performance in recent years. However, most existing deep SIQM evaluators are not specifically built for stereoscopic contents and consider little prior domain knowledge of the 3D human visual system (HVS) in network design. In this paper, we develop a Predictive Auto-encoDing Network (PAD-Net) for blind/No-Reference stereoscopic image quality measurement. In the first stage, inspired by the predictive coding theory that the cognition system tries to match bottom-up visual signal with top-down predictions, we adopt the encoder-decoder architecture to reconstruct the distorted inputs. Besides, motivated by the binocular rivalry phenomenon, we leverage the likelihood and prior maps generated from the predictive coding process in the Siamese framework for assisting SIQM. In the second stage, quality regression network is applied to the fusion image for acquiring the perceptual quality prediction. The performance of PAD-Net has been extensively evaluated on three benchmark databases and the superiority has been well validated on both symmetrically and asymmetrically distorted stereoscopic images under various distortion types.
\end{abstract}


%
\IEEEpeerreviewmaketitle

\section{Introduction}
%
%
%
%

\IEEEPARstart{S}{tereoscopy} is a technology that can create or enhance the illusion of depth by means of stereopsis for binocular vision. Stereoscopic image, as a popular image format, usually presents two offset images separately to the left and right eye of the observers \cite{background}. High-quality stereoscopic images increase the immersive feeling of consumers with the additional 3D depth perception and are essential for a wide scope of daily applications, e.g. 3D reconstruction \cite{xue2013matching, anchini2006comparison}, depth estimation \cite{rajagopalan2004depth, allison2009binocular} and object detection \cite{guo2014integrated}. However, through the process of acquisition, compression, transmission, display, etc. original reference stereoscopic images usually suffer from perceptual quality degradation caused by diverse distortion types and degrees \cite{de2008vector, angrisani2013internet, yue2018effective, jiang2020blind, jiang2020full}. Thus, to generate high-quality stereoscopic images and improve the user experience \cite{russo2005automatic, que2019exposure}, it is urgent to effectively measure the perceptual quality of stereoscopic images.

Compared with the conventional 2D image quality measurement (IQM) case, stereoscopic image quality measurement (SIQM) is more challenging owing to a wide variety of 2D and 3D influential factors, such as image spatial artifacts, depth perception, visual comfort, and so on \cite{series2012subjective}. These factors have different effects on evaluating the quality of experience (QoE) for stereoscopic images. Except for \cite{chen2017blind} that considers image distortion and depth perception quality simultaneously, existing research works mainly focus on modeling each individual factor for 3D QoE \cite{shao2016toward,zhou20163d,wang2016perceptual,chen2017visual}. In this paper, we aim to study the visually perceptual quality measurement of stereoscopic images.

Similar to 2D IQM, according to the availability of original reference stereoscopic images, SIQM models are typically divided into three categories: full-reference (FR) \cite{benoit2009quality,you2010perceptual,gorley2008stereoscopic,chen2013full,lin2014quality}, reduced-reference (RR) \cite{qi2015reduced,ma2016reorganized,ma2017reduced}, and blind/no-reference (NR) \cite{akhter2010no,sazzad2012objective,chen2013no,su2015oriented,oh2017blind,yang2019blind,zhou2019dual} SIQM metrics.

For FR SIQM algorithms, full information of the reference image is assumed to be exploited. The earliest FR IQM model investigated some off-the-shelf 2D IQM metrics, such as structural similarity index (SSIM) \cite{wang2004image}, universal quality index (UQI) \cite{wang2002universal}, C4 \cite{carnec2003image} and so on \cite{wang2005reduced}, to measure stereoscopic image quality \cite{campisi2007stereoscopic}. Further, disparity information was integrated into the 2D IQM metrics to predict the 3D image quality \cite{benoit2009quality,you2010perceptual}. Apart from incorporating depth clues, the binocular vision characteristics of the human visual system (HVS) were combined with 2D IQM algorithms. For example, Gorley and Holliman proposed a new stereo band limited contrast (SBLC) metric based on the HVS sensitivity to contrast changes in high frequency regions \cite{gorley2008stereoscopic}. Moreover, Chen \textit{et al.} \cite{chen2013full} proposed cyclopean images according to the binocular rivalry in the human eyes. In addition, Lin and Wu developed a binocular integration based computational model for evaluating the perceptual quality of stereoscopic images \cite{lin2014quality}.

As for RR SIQM approaches, only part of original non-distorted image data is available. Qi \textit{et al.} utilized binocular perceptual information (BPI) to perform RR SIQM \cite{qi2015reduced}. By characterizing the statistical properties of stereoscopic images in the reorganized discrete cosine transform (RDCT) domain, Ma \textit{et al.} presented the RR SIQM method \cite{ma2016reorganized}. Furthermore, a new RR SIQM metric based on natural scene statistics (NSS) and structural degradation was proposed \cite{ma2017reduced}.

\begin{figure*}[t]
	\centerline{\includegraphics[width=18cm]{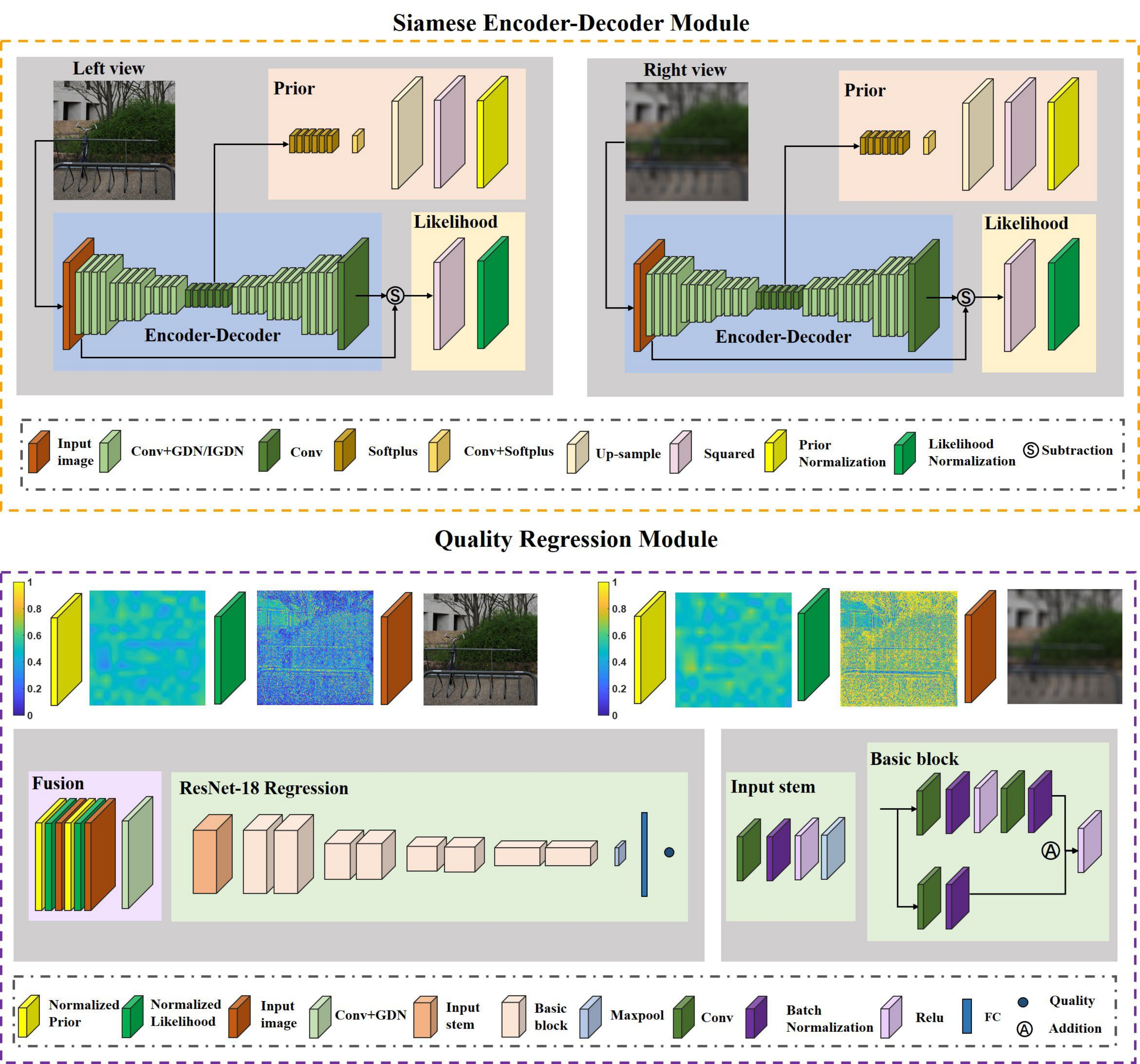}}
	\caption{The architecture of our proposed PAD-Net. It consists of a Siamese encoder-decoder module and a quality regression module. For the entire network, given paired distorted stereoscopic images and it will predict the perceptual quality score. In the Siamese encoder-decoder module, we calculate the prior and likelihood probability for each view image. Then, the left and right view images as well as their likelihood and prior probability maps are feed into the quality regression network for final score computation.}
	\centering
	\label{fig:fig1}
\end{figure*}

However, in most practical applications, the original pristine image cannot be accessible. Therefore, NR SIQM is inevitably required. Some research works have been studied about NR SIQM. Akhter \textit{et al.} extracted the local features of artifacts and disparity to evaluate the perceptual quality of stereoscopic images \cite{akhter2010no}. Sazzad \textit{et al.} also exploited perceptual differences of local features for NR SIQM \cite{sazzad2012objective}. However, these methods are distortion-specific NR SIQM approaches, which are only suitable for JPEG coded stereoscopic image pairs. Thus, several general-purpose NR SIQM metrics have emerged. Chen \textit{et al.} \cite{chen2013no} extracted 2D and 3D NSS features from stereoscopic image pairs. The shape-parameter features were regressed onto subjective quality scores by using the well-known support vector regression (SVR). Su \textit{et al.} proposed the stereoscopic/3D blind image naturalness quality (S3D-BLINQ) index by constructing a convergent cyclopean image and extracting bivariate and correlation NSS features in spatial and wavelet domains \cite{su2015oriented}.

With the development of deep learning techniques, deep neural networks (DNN) have achieved remarkable advantages for many image processing and computer vision tasks \cite{simonyan2014very,he2016deep,zhou2020blind}. It also brings improvement for the study on NR SIQM. Specifically, Shao \textit{et al.} proposed a blind deep quality evaluator for measuring stereoscopic image quality with monocular and binocular interactions based on deep belief network (DBN) \cite{shao2016toward}. The deep NR stereoscopic/3D image quality evaluator (DNR-S3DIQE) was proposed, which extracted local abstractions and then aggregated them into global features by employing the aggregation layer \cite{oh2017blind}. In addition, Yang \textit{et al.} took into account the deep perception map and binocular weight model along with the DBN to predict perceived stereoscopic image quality \cite{yang2019blind}. In \cite{yang2018blind}, a deep edge and color signal integrity evaluator (DECOSINE) was proposed based on the whole visual perception route from eyes to the frontal lobe. Besides, Zhou \textit{et al.} proposed a dual-stream interactive network called stereoscopic image quality assessment network (StereoQA-Net) for NR SIQM \cite{zhou2019dual}. However, the above-mentioned algorithms consider little prior domain knowledge of the 3D HVS, and thus having difficulty in accurately predicting the perceptual quality of stereoscopic images with various distortion types and levels.

Binocular vision is crucial to quality measurement for stereoscopic images, and can be mainly classified into three categories, namely binocular fusion, rivalry and suppression \cite{chen2017blind}. Firstly, if the retina regions of left and right eyes receive the same or similar visual contents, binocular fusion happens and the two views are integrated into one single and stable binocular perception \cite{howard1995binocular}. Secondly, binocular rivalry is a phenomenon in 3D vision, in which perception alternates between different views when two eyes see different scenes \cite{wang2015quality}. Thirdly, binocular suppression occurs since the HVS cannot tolerate binocular rivalry for a long time. During binocular suppression, one view may be inhibited by the other entirely \cite{chen2017blind}. Existing researches consider either one or multiple binocular vision mechanisms to assist SIQM and in this paper, we primarily focus on modeling binocular rivalry for SIQM. In conventional perspective, binocular rivalry is simulated by low-level competition between the input stimulus and it is related to the energy of the stimulus \cite{ohzawa1998mechanisms,levelt1965binocular}. Recently, more literatures try to explain binocular rivalry by the predictive coding theory \cite{dayan1998hierarchical,hohwy2008predictive}. It is a popular theory about how brain process sensing visual stimuli. According to predictive coding theory, the cognition system tries to match bottom-up visual signal with top-down predictions \cite{spratling2017review}. Different from the traditional statements that binocular rivalry is low-level inter-ocular competition in early visual cortex, binocular rivalry mechanism based on predictive coding theory (BRM-PC) stresses both on the low-level and high-level competition \cite{dayan1998hierarchical}. Moreover, the BRM-PC is the HVS guided and more inline with human cognition system \cite{hohwy2008predictive}. Therefore, we believe that introducing BRM-PC will be beneficial to SIQM.

In this paper, we propose a generic architecture called Predictive Auto-encoDing Network (PAD-Net), which is an end-to-end network for general-purpose NR SIQM. Our contributions of the proposed method are summarized as follows:

\begin{itemize}
\item We propose a biologically plausible and explicable predictive auto-encoding network through combining the network design with the binocular rivalry mechanism based on predictive coding theory \cite{hohwy2008predictive} which helps to explain the binocular rivalry phenomenon in 3D vision. Specifically, we adopt the encoder-decoder architecture to reconstruct the sensory input and further exploit the Siamese network to generate the corresponding likelihood as well as prior maps for modeling binocular rivalry in the cognition system. The source code of PAD-Net is available online for public research usage$\footnote{\url{http://staff.ustc.edu.cn/~chenzhibo/resources.html}}$.
\item We demonstrate that we can obtain the fusion information to reflect the perceived differences for symmetrically and asymmetrically distorted stereoscopic image pairs under various distortion types and levels.
\item Compared with state-of-the-art SIQM metrics, the proposed PAD-Net provides more precise quality estimation, especially for those stereopairs under asymmetric distortions thanks to the well consideration of binocular rivalry based on predictive coding theory.
\end{itemize}

The remainder sections of this paper are organized as follows. Section II explains the proposed Predictive Auto-encoDing Network (PAD-Net) for NR SIQM in details. Section III presents the experimental results and analysis. In Section IV, we conclude the paper with an outlook on the future work.

\section{Binocular Rivalry Mechanism Based on Predictive Coding Theory}
In this section, we will first describe the predictive coding theory and then introduce the binocular rivalry mechanism based on predictive coding theory (BRM-PC) in details.
\subsection{Predictive Coding Theory}
According to \cite{hume2003treatise}, the core task of brain is to represent the environmental causes of its sensory input. In other words, given a sensory input, the cognition system of human brain will predict the environmental cause, and gives a hypothesis. The final perceptual content is determined by the hypothesis that generates the best prediction \cite{hohwy2008predictive}.

\begin{figure}[h]
	\centerline{\includegraphics[width=9cm]{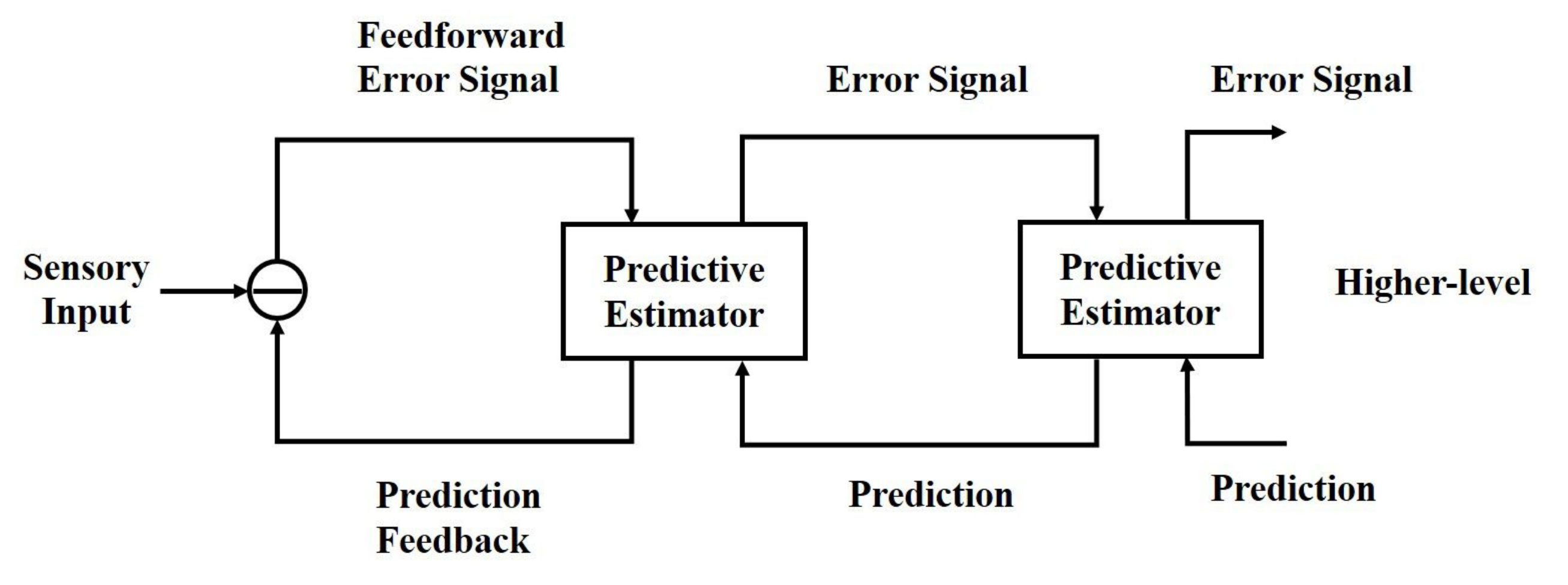}}
	\caption{Predictive coding theory based hierarchical representation of the sensory input \cite{rao1999predictive}. In this model, the feedback pathways carry the prediction from higher-level. In addition, the feedforward pathways return the prediction errors between the prediction and sensory input to update the prediction and get the best hypothesis.}
	\centering
	\label{fig:fig2}
\end{figure}

Numerous predictive coding models have been proposed, and the simplest is the linear predictive coding (LPC) in digital signal processing \cite{makhoul1975linear}. Then, the predictive coding theory has been applied to efficient encoding in the retina. Rao \textit{et al.} \cite{rao1999predictive} proposed a hierarchical model to represent the principle of human brain in Fig. \ref{fig:fig2}. The feedback pathways carry the prediction from higher-level and the feedforward pathways return the prediction errors between the prediction and sensory input to update the prediction and get the best hypothesis. The model mentioned here is for monocular vision, and it is the basis of binocular rivalry model.

\subsection{Binocular Rivalry Mechanism Based on Predictive Coding Theory}
According to \cite{wang2015quality}, binocular rivalry appears when left and right views are given different images, and the perception alternates between the two views. It is an important phenomenon when evaluating the quality of stereoscopic images due to the existence of asymmetrical distortion which means left and right views may suffer from different levels of distortion. Thus, employing binocular rivalry mechanism is beneficial to the improvement of stereoscopic image quality measurement.

\begin{figure}[ht]
	\centerline{\includegraphics[width=7.5cm]{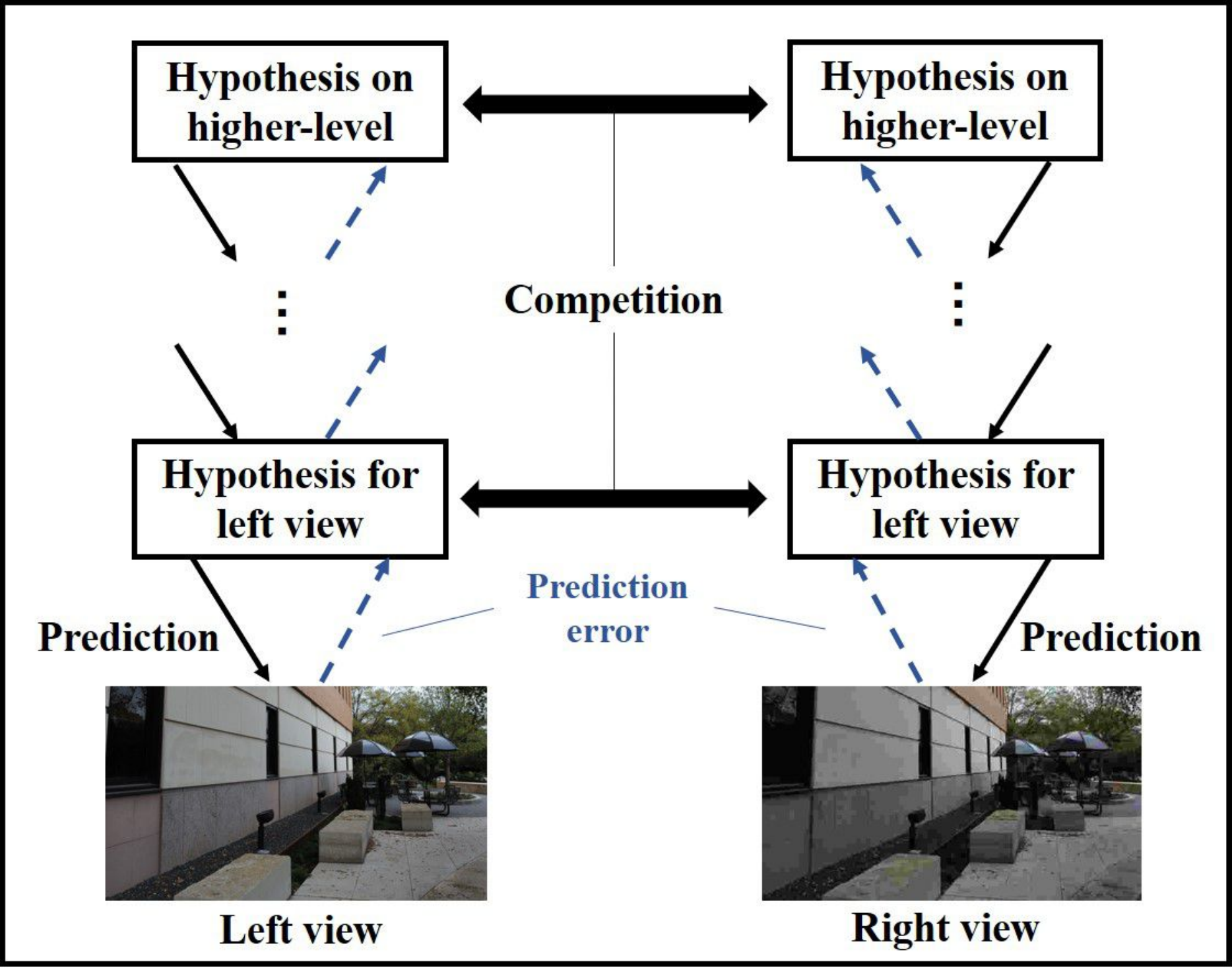}}
	\caption{ Simplified schematic of the binocular rivalry mechanism based on predictive coding theory \cite{hohwy2008predictive}. The black solid arrow is the top-down prediction from higher-level and the blue dotted arrow is the bottom-up error signals. The hypothesis for left and right view will compete with each other to generate the final perceptual content.}
	\centering
	\label{fig:fig3}
\end{figure}

Introducing predictive coding theory to explain binocular rivalry phenomenon has attracted greater attention from the community in recent years \cite{hohwy2008predictive}. Compared with the conventional perspective that believes binocular rivalry is low-level inter-ocular competition in early visual cortex, BRM-PC stresses both on the low-level and high-level competition \cite{leopold1996activity}. In this paper, we adopt the general theoretical framework in \cite{hohwy2008predictive,chen2020stereoscopic} for SIQM. On the basis of BRM-PC, we perceive the content since the corresponding hypothesis has the higher posterior probability from the Bayesian perspective \cite{friston2002functional,kersten2004object}. As shown in Fig. \ref{fig:fig3}, given a stereoscopic image, our brain normally first determines a hypothesis that can best explain the corresponding stimulus. The perceptual inference depends on the likelihood as well as the prior probability of the hypotheses. Note that we are not going to calculate real probabilities in the proposed method, but just try to model the likelihood and prior with a quantity that has similar physical meaning in the BRM-PC described in \cite{hohwy2008predictive}. The likelihood $p(\bm{I}|\bm{H})$ is about how well the hypothesis $\bm{H}$ predicts the input $\bm{I}$. Therefore, we apply the error between the sensory input and the prediction to compute the likelihood map. Specifically, small error denotes large likelihood. Besides, the prior $p(\bm{H})$ is related to empirical knowledge (shape, material, lighting) which comes from hypothesis on higher-level \cite{kersten2004object,ulyanov2018deep} and is about how probable the hypothesis is. Thus, we generate the prior map with the high-level features. Then, the hypotheses for left and right views will compete with each other according to the posterior probability $p(\bm{H}|\bm{I})$ computing from prior and likelihood.

According to the above analysis, we can conclude that the likelihood and prior are important to obtain perceptual inference during binocular rivalry. The details about how BRM-PC related to our proposed model is presented in the next section.

\section{Proposed Method}
Inspired by the BRM-PC \cite{hohwy2008predictive}, we design the no-reference PAD-Net to automatically predict the perceptual quality of stereoscopic images. The proposed PAD-Net is a Siamese based end-to-end network, including auto predictive coding and quality regression modules. Given paired distorted 3D images, they are divided into sub-images with size 256x256 first, then the quality scores of these sub-images are estimated through PAD-Net and aggregated as a final one as done in \cite{ma2017end}. We pre-train the proposed ‘auto-encoder’ and quality regression module on Waterloo Exploration Database \cite{ma2016waterloo} and LIVE 2D Database \cite{sheikh2006statistical}, respectively. After that, the entire network is jointly optimized using 3D IQM databases \cite{moorthy2013subjective,chen2013full} to generate more accurate predictions.

In this section, the architecture of our proposed PAD-Net is described first as shown in Fig. \ref{fig:fig1}. Then, we introduce the Siamese encoder-decoder module and quality regression module in details. Finally, the training and testing methods of our PAD-Net are presented.

\subsection{Overview}
The architecture of the PAD-Net is depicted in Fig. \ref{fig:fig1}, which contains a Siamese encoder-decoder module and a quality regression module. The Siamese encoder-decoder module represents the processing procedure of left and right view images. It is inspired by the predictive coding theory that the human brain tries to match bottom-up visual stimuli with top-down predictions \cite{spratling2017review}. In Section II, an empirical Bayesian framework based on BRM-PC is introduced and the rivalry dominance is calculated from the likelihood and prior probability \cite{tong2006neural}. Correspondingly, the likelihood and prior are inferred through the Siamese encoder-decoder in our proposed PAD-Net. In Fig. \ref{fig:fig1}, the normalized prior and likelihood maps are visualized. The higher the value (from blue to yellow) is, the larger the probability. Fig. \ref{fig:fig1} gives an example of asymmetrically blurred image, the left view is undamaged while the right view is blurred. Observed from the prior map, the discrepancy between normalized left and right view prior maps is not significant since we can still recognize the scene for both views. For image blur, it is reported in \cite{wang2015quality} that 3D image quality is more affected by the high-quality view. As indicated in the likelihood map, strong edges in the undistorted view tend to gain more probability (yellow color) in binocular rivalry, which demonstrates that subjective judgments are more affected by structural information \cite{wang2004image}.

For each branch of the Siamese encoder-decoder, four times down convolution and up convolution with a stride of 2 pixels are performed to reconstruct the distorted images. Non-linear activation function layers follow the first three down convolution and up convolution layers to enhance the representation ability of our neural network. We square the error between the input and reconstructed image as the residual map and utilize this map for likelihood calculation. Moreover, the high-level features are convolved with a stride of one pixel to change the channel size and upscaled to the same resolution as the input image. In the PAD-Net, we use the reconstruction error to obtain likelihood probability map and the high-level representation after four times down convolution to generate prior probability map. The reasons will be given in Section III B.

Once we get the left and right view images as well as their likelihood and prior probability maps, we can put them into the quality regression module and compute the final quality score. It can be seen from the bottom of Fig. \ref{fig:fig1} that the quality regression module is composed of 1) fusion for input images and probability maps, 2) ResNet-18 without last two layers as feature extractor and 3) the final max pooling and fully connected layer. Note that fusion in Fig. \ref{fig:fig1} includes a one stride convolution layer and activation function layer to match the input size with ResNet-18. The output of the proposed PAD-Net is the predicted quality score of the input stereoscopic image.

\subsection{Siamese Encoder-Decoder Module}

The Siamese encoder-decoder module is inspired by the BRM-PC, which is the HVS guided and able to generate more reliable and interpretable results. Based on the predictive coding process in BRM-PC, we adopt the encoder-decoder network structure for image compression from \cite{balle2016end,balle2018variational}, which is a hierarchical structure. In Fig. \ref{fig:fig2}, prediction is adjusted with residual errors to obtain better hypothesis, thus, we employ the squared error between predicted (decoded) image and input image \cite{rao1999predictive} as our loss function $l_1$ to pre-train the encoder-decoder network as follows:
\begin{equation}\label{1}
l_1 = \frac{1}{{M \!\times \!N \!\times \!C \!\times \!K}}\sum\limits_{x,y,c,k} {{{( \bm{I}^{(k)}(x,y,c) - \bm{\widehat I}^{(k)}(x,y,c))}^2}},
\end{equation}
\begin{equation}\label{2}
\bm{\widehat I}^{(k)} = f_1(\bm{I}^{(k)};\bm{w}_1),
\end{equation}
\begin{equation}\label{3}
\bm{w}_{1}^{'} = \arg \mathop {\min }\limits_{\bm{w}_1} l_1,
\end{equation}
where $\bm{I}^{(k)}$ and $\bm{\widehat I}^{(k)}$ represent the $k$-th input and predicted image. $M$, $N$ and $C$ are the width, height and channel of $\bm{I}^{(k)}$ and $\bm{\widehat I}^{(k)}$. $K$ is the batch size of a mini-batch training data. $\bm{\widehat I}^{(k)}$ is estimated via the encoder-decoder network $f_1$ with weight $\bm{w}_1$. $\bm{w}_{1}^{'}$ is the weight for encoder-decoder after loss minimization. During the training stage, the output of decoder can be viewed as the feedback prediction in Fig. \ref{fig:fig2}, and the loss is defined as the prediction error between distorted and reconstructed images which can be assumed as the feedforward error signal in Fig. \ref{fig:fig2}. Then, we use the gradient descent algorithm to update $\bm{w}_1$ and generate better prediction. Finally, the prediction error will converge and reach a stable value. The decoded image would not change greatly. Thus, the updating policy of the encoder-decoder network is similar to the prediction coding theory of human brain. Note that the encoder-decoder network is pre-trained on the Waterloo Exploration Database \cite{ma2016waterloo}, therefore the training data includes both reference and distorted images.

\begin{table}[ht]
\begin{center}
\captionsetup{justification=centering}
\caption{\textsc{Detailed Configurations of the Encoder-decoder Network.}}
\label{table1}
\begin{threeparttable}
\scalebox{0.85}{
\begin{tabular}{@{}c|cccc@{}}
\toprule
 & Layer & \begin{tabular}[c]{@{}c@{}}Input Shape\\ $C \times H \times W$\end{tabular} & \begin{tabular}[c]{@{}c@{}}Output Shape\\ $C \times H \times W$\end{tabular} & \begin{tabular}[c]{@{}c@{}}Kernel size/\\ Stride/Padding\end{tabular} \\ \midrule
\multirow{4}{*}{Encoder} & Conv1+GDN1 & 3, 256, 256 & 128, 128, 128 & 5x5/2/2 \\
 & Conv2+GDN2 & 128, 128, 128 & 128, 64, 64 & 5x5/2/2 \\
 & Conv3+GDN3 & 128, 64, 64 & 128, 32, 32 & 5x5/2/2 \\
 & Conv4 & 128, 32, 32 & 192, 16, 16 & 5x5/2/2 \\ \midrule
\multirow{4}{*}{Decoder} & Unconv1+IGDN1 & 192, 16, 16 & 128, 32, 32 & 5x5/2/2 \\
 & Unconv2+IGDN2 & 128, 32, 32 & 128, 64, 64 & 5x5/2/2 \\
 & Unconv3+IGDN3 & 128, 64, 64 & 128, 128, 128 & 5x5/2/2 \\
 & Unconv4 & 128, 128, 128 & 3, 256, 256 & 5x5/2/2 \\ \bottomrule
\end{tabular}}
 \begin{tablenotes}
        \footnotesize
        \item[1] Conv: Convolution
	   \item[2] Unconv: Fractionally-strided convolution
        \item[3] GDN: Generalized divisive normalization
	   \item[4] IGDN: Inverse generalized divisive normalization
	   \item[5] $C \times H \times W$: Channel $\times$ Height $\times$ Width
      \end{tablenotes}
\end{threeparttable}
\end{center}
\end{table}

The structure of the encoder-decoder network is listed in Table \ref{table1}. It is composed of convolution and fractionally-strided convolution layers \cite{radford2015unsupervised}, generalized divisive normalization (GDN) transform and inverse generalized divisive normalization (IGDN). GDN is inspired by the neuron models in biological visual system and proves to be effective in density estimation \cite{balle2015density}, image compression \cite{balle2018variational} and image quality assessment \cite{ma2017end}. The GDN and IGDN operations are given by:
\begin{equation}\label{4}
{\bm{y}_i}(m,n) = \frac{{\bm{x}_i(m,n)}}{{({\beta _i} + \sum\nolimits_j {{\gamma _{ij}}\bm{x}_j{{(m,n)}^2}{)^{\frac{1}{2}}}} }},
\end{equation}
\begin{equation}\label{5}
\bm{\widehat x}_i(m,n) = \bm{\widehat y}_i(m,n) \cdot ({\widehat \beta _i} + \sum\nolimits_j {{{\widehat \gamma }_{ij}}\bm{x}_j{{(m,n)}^2}{)^{\frac{1}{2}}}},
\end{equation}
where $\bm{x}_i$ means the $i$-th channel of $\bm{x}$ and it is the input of GDN transform. $\bm{y}_i$ is the $i$-th channel of normalized activation feature map and it is the output of GDN operation. Moreover, $\bm{\beta}$ and $\bm{\gamma}$ are the parameters to be updated in GDN function. Likewise, $\bm{\widehat x}_i$, $\bm{\widehat y}_i$, $\bm{\widehat \beta} $ and $\bm{\widehat \gamma}$ share the same meaning as $\bm{x}_i$, $\bm{y}_i$, $\bm{\beta} $ and $\bm{\gamma}$ for IGDN transform. The goal of our encoder-decoder module is to reconstruct the input sensory instead of compressing it. Therefore, we remove the quantization step in \cite{balle2018variational} for better reconstruction.

By now, we can get the high-level causes (high-level features) and reconstructed image from the input sensory in the proposed Siamese encoder-decoder module. Then, according to the BRM-PC \cite{hohwy2008predictive}, the left and right views will compete with each other to obtain the best hypothesis which is related to prior and likelihood. To be specific, the likelihood is about how well the hypothesis predicts the input and the prior is about how probable the hypothesis is and concern with empirical knowledge \cite{kersten2004object}. Corresponding to the physical meaning of likelihood and prior, we first obtain the $k$-th squared residual error map $\bm{E}^{(k)}$ in the mini-batch of a size of $K$ to calculate likelihood as follows:
\begin{equation}\label{6}
\bm{E}^{(k)} = \frac{1}{C}\sum\nolimits_c {{{(\bm{I}_c^{(k)} - \bm{\widehat I}_c^{(k)})}^2}},
\end{equation}
where $\bm{I}_c^{(k)}$ and $\bm{\widehat I}_c^{(k)}$ denote the $c$-th channel of the input and predicted image. $C$ equals 3 in Eq. \ref{6}. The likelihood is used to measure the similarity between $\bm{I}_c^{(k)}$ and $\bm{\widehat I}_c^{(k)}$ and it is inversely proportional to errors. The training stage of encoder-decoder network can be regarded as the procedure of prediction error minimization.

In the Siamese encoder-decoder module, the prior is modeled with the high-level representation of the sensory input \cite{ulyanov2018deep} since the prior comes from high-levels in the cognition system, as assumed in empirical Bayes \cite{hohwy2008predictive,summerfield2005mistaking}. Thus, the high-level features are utilized to generate the $k$-th prior map $\bm{P}^{(k)}$. Before being fed into the quality regression module, the squared error map $\bm{E}^{(k)}$ and the prior map $\bm{P}^{(k)}$ are normalized between left and right view as follows:
\begin{equation}\label{7}
\bm{P}_{nl}^{(k)} = \frac{{\bm{P}_{l}^{(k)}}}{{\bm{P}_{l}^{(k)} + \bm{P}_{r}^{(k)}}} \quad and \quad  \bm{P}_{nr}^{(k)} = \frac{{\bm{P}_{r}^{(k)}}}{{\bm{P}_{l}^{(k)} + \bm{P}_{r}^{(k)}}},
\end{equation}
\begin{equation}\label{8}
\bm{L}_{nl}^{(k)} = \frac{{\bm{E}_{r}^{(k)}}}{{\bm{E}_{l}^{(k)} + \bm{E}_{r}^{(k)}}} \quad and \quad  \bm{L}_{nr}^{(k)} = \frac{{\bm{E}_{l}^{(k)}}}{{\bm{E}_{l}^{(k)} + \bm{E}_{r}^{(k)}}},
\end{equation}
where $\bm{P}_{l}^{(k)}$, $\bm{P}_{r}^{(k)}$, $\bm{E}_{l}^{(k)}$ and $\bm{E}_{r}^{(k)}$ are the prior and error maps for the $k$-th left view and right view images in a mini-batch. $\bm{P}_{nl}^{(k)}$, $\bm{P}_{nr}^{(k)}$, $\bm{L}_{nl}^{(k)}$ and $\bm{L}_{nr}^{(k)}$ indicate the normalized prior and likelihood probability maps. Note that the error is opposite to the likelihood, that is to say, if the error is large, the likelihood will be small. For example, when computing the likelihood map for left view, the error map of right view is adopted and vice versa.

\begin{table}[ht]
\begin{center}
\captionsetup{justification=centering}
\caption{\textsc{Detailed Configurations of the Prior and Likelihood Generation Network.}}
\label{table2}
\begin{threeparttable}
\scalebox{0.85}{
\begin{tabular}{@{}c|cccc@{}}
\toprule
 & Layer & \begin{tabular}[c]{@{}c@{}}Input Shape\\ $C \times H \times W$\end{tabular} & \begin{tabular}[c]{@{}c@{}}Output Shape\\ $C \times H \times W$\end{tabular} & \begin{tabular}[c]{@{}c@{}}Kernel size/\\ Stride/Padding\end{tabular} \\ \midrule
\multirow{5}{*}{Prior} & Softplus4 & 192, 16, 16 & 192, 16, 16 & - \\
 & Conv5+Softplus5 & 192, 16, 16 & 1, 16, 16 & 1x1/1/0 \\
 & Up-sample6 & 1, 16, 16 & 1, 256, 256 & - \\
 & Square7a & 1, 256, 256 & 1, 256, 256 & - \\
 & Normlization8a & 1, 256, 256 & 1, 256, 256 & - \\ \midrule
\multirow{2}{*}{Likelihood} & Square7b & 1, 256, 256 & 1, 256, 256 & - \\
 & Normlization8b & 1, 256, 256 & 1, 256, 256 & - \\ \bottomrule
\end{tabular}}
 \begin{tablenotes}
        \footnotesize
	   \item[2] Normlization: Normalization between left view and right view
      \end{tablenotes}
\end{threeparttable}
\end{center}
\end{table}

The detailed structure of prior and likelihood creation is given in Table \ref{table2}. We employ the Softplus activation function \cite{glorot2011deep} in prior generation to avoid square negative values to positive values. It is defined as:
\begin{equation}\label{9}
s(x) = \log (1 + {e^x}).
\end{equation}
The Softplus function can be regarded as the smoothing version of ReLU function which is similar to the way cerebral neurons being activated \cite{nair2010rectified}.

\subsection{Quality Regression Module}

Based on the distorted stereoscopic images as well as the obtained prior and likelihood probability maps from the Siamese encoder-decoder module, we fuse them as a 3-channel feature map and further feed the 3-channel feature map into the ResNet-18 quality regression network to extract discriminative features for quality estimation. ResNet-18 is chosen for its excellent ability of feature extraction \cite{he2016deep,he2016identity}. The last two layers including average pooling and fully connected layer are removed for regressing the feature map into a quality score. Table \ref{table3} illustrates the architecture of quality regression network. The input stem and basic block of the ResNet-18 structure are shown in Fig. \ref{fig:fig1}.

\begin{table}[ht]
\begin{center}
\captionsetup{justification=centering}
\caption{\textsc{Detailed Configurations of the Quality Regression Network.}}
\label{table3}
\begin{threeparttable}
\scalebox{0.85}{
\begin{tabular}{@{}c|cccc@{}}
\toprule
 & Layer & \begin{tabular}[c]{@{}c@{}}Input Shape\\ $C \times H \times W$\end{tabular} & \begin{tabular}[c]{@{}c@{}}Output Shape\\ $C \times H \times W$\end{tabular} & \begin{tabular}[c]{@{}c@{}}Kernel size/\\Stride/Padding\end{tabular} \\ \midrule
Fusion & Conv9+GDN9 & 10, 256, 256 & 3, 256, 256 & 1x1/1/0 \\ \midrule
\multirow{3}{*}{Regression} & ResNet-18 & 3, 256, 256 & 512,8,8 & - \\
 & Maxpool10 & 512,8,8 & 512,1,1 & 8x8/8/0 \\
 & Fc11 & 512 & 1 & -  \\ \bottomrule
\end{tabular}}
 \begin{tablenotes}
        \footnotesize
	   \item[3] Fc: Fully connected layer
      \end{tablenotes}
\end{threeparttable}
\end{center}
\end{table}

\subsection{Training and Testing}
Owing to the limited size of available 3D image quality measurement database, we train the PAD-Net on the sub-image pairs with the resolution of $256\times256$, the MOS value for the entire image is assumed as the quality scores for several sub-images as done in \cite{ma2017end}. Thus, sub-images coming from the same test image share the same labels. Moreover, transfer learning is adopted to solve the problem of lacking labeled data and enhance the prediction accuracy of the network \cite{pan2009survey}.

In blind stereoscopic image quality measurement, it is difficult to predict the MOS value precisely \cite{ma2017end}. Therefore, we divide the training stage into three steps: 1) pre-training of encoder-decoder on the pristine and distorted 2D images of Waterloo Exploration Database \cite{ma2016waterloo}; 2)  pre-training ResNet-18 regression on the pristine and distorted 2D images of LIVE 2D Database \cite{sheikh2006statistical}; 3) joint optimization on the 3D IQM database.

Firstly, encoder-decoder is trained to minimize the difference between predicted and input images which is described in Section III B. Then, we get the weight $\bm{w}_1$ for the Siamese encoder-decoder as follows:
\begin{equation}\label{10}
\bm{w}_{1}^{'} = \arg \mathop {\min }\limits_{\bm{w}_1} {l_1}.
\end{equation}

Secondly, we utilize the original and distorted 2D images along with the associated MOS scores to pre-train the ResNet-18 regression network. It is aimed to map the 2D image into a quality score. In addition, ResNet-18 with pre-trained weight on ImageNet is adopted for better initialization. Then, the loss function $l_2$ for second step pre-training is defined as:
\begin{equation}\label{11}
{l_2} = \frac{1}{K}\sum\nolimits_k {{{(q_{2d}^{(k)} - \widehat q_{2d}^{(k)})}^2}},
\end{equation}
\begin{equation}\label{12}
\widehat q_{2d}^{(k)}{\rm{ = }}{f_2}({\bm{I}^{(k)}};{\bm{w}_2}),
\end{equation}
where $q^{(k)}_{2d}$ and  $\widehat q^{(k)}_{2d}$ indicate the real MOS and predicted score for the $k$-th input 2D sub-image $\bm{I}^{(k)}$ in a mini-batch. The weight $\bm{w}_2$ for the ResNet-18 regression network $f_2$ is updated by minimizing $l_2$ as follows:
\begin{equation}\label{13}
\bm{w}_{2}^{'} = \arg \mathop {\min }\limits_{\bm{w}_2} {l_2}.
\end{equation}

Finally, the Siamese encoder-decoder and quality regression module are jointly optimized using stereo image pairs. Since the ultimate purpose of PAD-Net is to estimate the perceptual quality of 3D images, we again adopt the $l_2$-norm between the subjective MOS value $\bm{q}_{3d}$ and predicted score $\bm{\widehat q}_{3d} $ as loss function:
\begin{equation}\label{14}
{l_3} = \frac{1}{K}\sum\nolimits_k {{{(q_{3d}^{(k)} - \widehat q_{3d}^{(k)})}^2}},
\end{equation}
\begin{equation}\label{15}
\widehat q_{3d}^{(k)}{\rm{ = }}f(\bm{I}_l^{(k)},\bm{I}_r^{(k)};{\bm{w}_1},{\bm{w}_2},{\bm{w}_3}),
\end{equation}
where $\bm{I}^{(k)}_l$ and  $\bm{I}^{(k)}_r$ represent  the input 3D sub-image pairs. $f$ indicates the PAD-Net with encoder-decoder weight $\bm{w}_1$, ResNet-18 regression weight $\bm{w}_2$ and weight $\bm{w}_3$ which is trained from scratch. Specifically, $\bm{w}_3$ includes the parameters in prior generation part of Siamese encoder-decoder module and fusion part of quality regression module. At the joint optimization step, $(\bm{w}_1, \bm{w}_2)$ are initialized with pre-trained weight $(\bm{w}_1^{'}, \bm{w}_2^{'})$ and updated with $\bm{w}_3$ through final loss minimization:
\begin{equation}\label{16}
\bm{w}_1^*,\bm{w}_2^*,\bm{w}_3^* = \arg \mathop {\min }\limits_{{\bm{w}_1},{\bm{w}_2},{\bm{w}_3}} l_3.
\end{equation}

In the testing stage, the stereo image is divided into sub-image pairs with a stride of $U$ to cover the whole content. The predicted qualities of all sub-image pairs are averaged to compute the final perceptual quality score.

\section{Experimental Results and Analysis}
In this section, we first introduce the databases and performance measures used in our experiment. Then, the experimental results of the proposed PAD-Net on the entire LIVE databases and individual distortion type are illustrated. Meanwhile, the visualization results are provided for better explanation. Finally, we conduct the ablation study to verify the effectiveness of each component in our model and measure the computation complexity.

\subsection{Databases and Performance Measures}
Three benchmark stereoscopic image quality measurement databases are used in our experiment including LIVE Phase I \cite{moorthy2013subjective}, LIVE Phase II \cite{chen2013full,chen2013no} and Waterloo IVC Phase I \cite{wang2015quality}.

\textbf{LIVE Phase I \cite{moorthy2013subjective}:} This database contains 20 original and 365 symmetrically distorted stereo image pairs. Five distortion types are included in this database, namely JPEG2000 compression (JP2K), JPEG compression (JPEG), additive white noise (WN), Gaussian blur (BLUR) and Raleigh fast fading channel distortion (FF). Subjective differential mean opinion score (DMOS) is provided for each degraded stereo image. Higher DMOS value means lower visual quality.

\textbf{LIVE Phase II \cite{chen2013full,chen2013no}:} It includes 120 symmetrically and 240 asymmetrically distorted stereopairs derived from 8 reference images. This database contains the same distortion types as LIVE Phase I. For each distortion type, the pristine image pair is degraded to 3 symmetrically and 6 asymmetrically image pairs. Subjective scores are also recorded in the form of DMOS.

\textbf{Waterloo IVC Phase I \cite{wang2015quality}:} This database originates from 6 pristine stereoscopic image pairs. The reference image is altered by three types of distortions, namely WN, BLUR, and JPEG. Altogether, there are totally 78 symmetrically and 252 asymmetrically distorted stereopairs. Subjective mean opinion score (MOS) and individual scores are provided for each stereoscopic image in this database, while higher MOS value means better visual quality.

\textbf{Performance Measure:} Three commonly used criteria \cite{video2003final} are utilized in our experiment for performance evaluation, including Spearman's rank order correlation coefficient (SROCC), Pearson's linear correlation coefficient (PLCC) and root mean squared error (RMSE). SROCC is a non-parametric measure and independent of monotonic mapping. PLCC and RMSE evaluate the prediction accuracy. Higher SROCC, PLCC and lower RMSE indicate better correlation with human judgements. Before calculating PLCC and RMSE, a five-parameter logistic function \cite{sheikh2006statistical} is applied to maximize the correlation between subjective ratings and objective metrics.

One of the main issues of PLCC and SROCC is that they neglect the uncertainty of the subjective scores \cite{krasula2016accuracy}. Thus, we also employ the Krasula methodology \cite{krasula2016accuracy}, which could be used to better assess the capabilities of objective metrics by considering the statistical significance of the subjective scores and getting rid of the mapping functions. The basic idea of this model is to determine the reliability of objective models by checking whether they are capable of well 1) distinguishing the significantly different stimuli from the similar ones, and 2) indicating whether one stimulus are of better/worse quality than the other. To this end, in the  Krasula framework,  pairs of stimuli are selected from the database to compute the area under ROC curve of the ‘Different vs. Similar’ category (AUC-DS), area under ROC curve of the ‘Better vs.Worse’ category (AUC-BW), and percentage of correct classification (CC). Higher AUC-DS and AUC-BW mean more capability to indicate different/similar and better/worse pairs. Higher CC represents better prediction accuracy. Please refer to \cite{krasula2016accuracy} for more details.

\subsection{Performance Evaluation}
\begin{figure}[htbp]
  \centerline{\includegraphics[width=9cm]{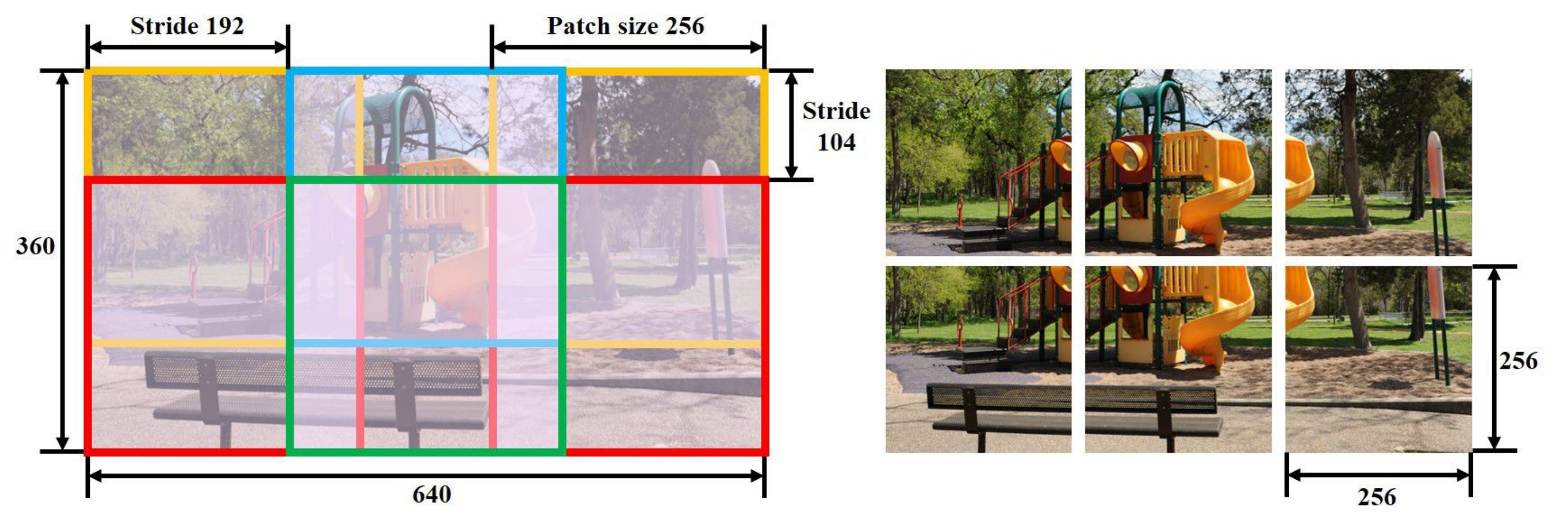}}
  \caption{Crop in a slight overlapping manner during the testing stage.}
  \centering
\label{fig:fig4}
\end{figure}

In the experiment, the distorted stereo pairs are randomly split into 80\% training set and 20\% testing set according to \cite{zhou2019dual}. We adopt the Adam algorithm in the pre-training and joint optimization step. During pre-training, the learning rate $\alpha$ is set as $10^{-4}$ and lowered by a factor of 10 every 50 epochs. The pre-trained weights are obtained after 100 epochs. Since encoder-decoder should retain its function, the learning rate $\alpha_1$ for encoder-decoder weight $\bm{w}_1$ is set as $10^{-5}$ to avoid drastic change when conducting joint optimization. Moreover, the learning rate $\alpha_2$ for ResNet-18 regression $\bm{w}_2$ is set as half of $\alpha_3$ for $\bm{w}_3$. $\alpha_3$ is initialized as $10^{-3}$ and scaled by 0.25 every 50 epochs. The learning rate remains unchanged after 200 epochs. We apply data augmentation by randomly cropping, horizontal and vertical flipping in the training stage \cite{mikolajczyk2018data}. The results are obtained after 300 epochs. During testing, the stride U is set as 192 for width and 104 for height in a slight overlapping manner to cover the whole resolution in LIVE databases as shown in Fig. \ref{fig:fig4}.

We compare the proposed PAD-Net with several classic FR, RR and NR SIQM metrics on the LIVE Phase I and II database. The competing FR and RR models include Gorley's method \cite{gorley2008stereoscopic} , You's method \cite{you2010perceptual}, Benoit's method \cite{benoit2009quality}, Lin's method \cite{lin2014quality}, Cyclopean MS-SSIM \cite{chen2013full}, RR-BPI \cite{qi2015reduced}, RR-RDCT \cite{ma2016reorganized} and Ma's method \cite{ma2017reduced}. For NR metrics, some hand-crafted features based algorithms including Akhter's method \cite{akhter2010no}, Sazzad's method  \cite{sazzad2012objective}, Chen's method \cite{chen2013no}, S3D-BLINQ \cite{su2015oriented}, DECOSINE \cite{yang2018blind} and deep neural network based models including Shao's method \cite{shao2016toward}, CNN \cite{kang2014convolutional}, DNR-S3DIQE \cite{oh2017blind}, DBN \cite{yang2019blind}, StereoQA-Net \cite{zhou2019dual} are considered in the performance comparison. Note that CNN \cite{kang2014convolutional} is computed for left and right view images separately and then average the scores for both views. The SROCC, PLCC and RMSE performance for the above metrics and proposed PAD-Net are listed in Table \ref{table4} where the best results are highlighted in bold. It could be observed from the table that the proposed method outperforms state-of-the-art SIQM metrics, especially on LIVE Phase II database. Since there are more asymmetrically distorted images in LIVE II, the proposed PAD-Net is more effective for the challenging asymmetric distortion which will be explained in Section III D.

\begin{table}[ht]
	\begin{center}
		\captionsetup{justification=centering}
		\caption{\textsc{Overall Performance Comparison on LIVE Phase I and II Databases. The Best Performing Results are Highlighted in Bold.}}
		\label{table4}
		\scalebox{0.75}{
			\begin{tabular}{@{}c|c|ccc|ccc@{}}
				\toprule
				&  & \multicolumn{3}{c|}{LIVE Phase I} & \multicolumn{3}{c}{LIVE Phase II} \\ \midrule
				Type & Metrics & SROCC & PLCC & RMSE & SROCC & PLCC & RMSE \\ \midrule
				\multirow{5}{*}{FR} & Gorley \cite{gorley2008stereoscopic} & 0.142 & 0.451 & 14.635 & 0.146 & 0.515 & 9.675 \\
				& You \cite{you2010perceptual} & 0.878 & 0.881 & 7.746 & 0.786 & 0.800 & 6.772 \\
				& Benoit \cite{benoit2009quality} & 0.899 & 0.902 & 7.061 & 0.728 & 0.748 & 7.490 \\
				& Lin \cite{lin2014quality} & 0.856 & 0.784 & - & 0.638 & 0.642 & - \\
				& Cyclopean MS-SSIM \cite{chen2013full} & 0.916 & 0.917 & 6.533 & 0.889 & 0.900 & 4.987 \\ \midrule
				\multirow{3}{*}{RR} & RR-BPI \cite{qi2015reduced} & - & - & - & 0.867 & 0.915 & 4.409 \\
				& RR-RDCT \cite{ma2016reorganized} & 0.905 & 0.906 & 6.954 & 0.809 & 0.843 & 6.069 \\
				& Ma \cite{ma2017reduced} & 0.929 & 0.930 & 6.024 & 0.918 & 0.921 & 4.390 \\ \midrule
				\multirow{11}{*}{NR} & Akhter \cite{akhter2010no} & 0.383 & 0.626 & 14.827 & 0.543 & 0.568 & 9.294 \\
				& Sazzad \cite{sazzad2012objective} & 0.624 & 0.624 & - & 0.648 & 0.669 & - \\
				& Chen \cite{chen2013no} & 0.891 & 0.895 & 7.247 & 0.880 & 0.895 & 5.102 \\
				& S3D-BLINQ \cite{su2015oriented} & - & - & - & 0.905 & 0.913 & 4.657 \\
				& Shao \cite{shao2016toward} & 0.945 & 0.957 & - & 0.911 & 0.927 & - \\
				& CNN \cite{kang2014convolutional} & 0.896 & 0.933 & 5.948 & 0.633 & 0.634 & 8.632 \\
				& DNR-S3DIQE \cite{oh2017blind} & 0.935 & 0.943 & - & 0.871 & 0.863 & - \\
				& DBN \cite{yang2019blind} & 0.944 & 0.956 & 4.917 & 0.921 & 0.934 & 4.005 \\
				& DECOSINE \cite{yang2018blind} & 0.953 & 0.962 & - & 0.941 & 0.950 & - \\
				& StereoQA-Net \cite{zhou2019dual} & 0.965 & 0.973 & 4.711 & 0.947 & 0.957 & 3.270 \\
				& Proposed PAD-Net & \textbf{0.973} & \textbf{0.975} & \textbf{3.514} & \textbf{0.967} & \textbf{0.975} & \textbf{2.446} \\ \bottomrule
		\end{tabular}}
	\end{center}
\end{table}

\begin{table}[ht]
	\begin{center}
		\captionsetup{justification=centering}
		\caption{\textsc{SROCC, PLCC, RMSE and Krasula Performance Evaluation on the Waterloo IVC Phase I Database. The Best Performing Results are Highlighted in Bold.}}
		\label{table13}
		\setlength{\tabcolsep}{5mm}{
		\scalebox{1}{
			\begin{tabular}{@{}c|ccc@{}}
				\toprule
				Metrics          & SROCC          & PLCC           & RMSE           \\ \midrule
				StereoQA-Net     & 0.955          & 0.970          & 4.350          \\
				Proposed PAD-Net & \textbf{0.974} & \textbf{0.979} & \textbf{3.696} \\ \midrule
				Metrics          & AUC\_DS        & AUC\_BW        & CC             \\ \midrule
				StereoQA-Net     & 0.894          & 0.998          & 0.979          \\
				Proposed PAD-Net & \textbf{0.925} & \textbf{0.999} & \textbf{0.993} \\ \bottomrule
		\end{tabular}}}
	\end{center}
\end{table}

To employ the Krasula methodology \cite{krasula2016accuracy}, significance analysis of the subjective scores are required. Among the three considered 3D IQM databases \cite{moorthy2013subjective,chen2013full,chen2013no,wang2015quality}, only the Waterloo IVC Phase I \cite{wang2015quality} is equipped with individual scores. Furthermore, the excusable/source code of most of the state-of-the-art  NR 3D metrics are not released. Therefore, we could only conduct the Krasula analysis on the Waterloo IVC Phase I dataset and compare the proposed PAD-Net with StereoQA-Net \cite{zhou2019dual},  which obtains the best performance on both  LIVE Phase I \cite{moorthy2013subjective}  and LIVE Phase II \cite{chen2013full,chen2013no} databases. Table \ref{table13} lists the results of SROCC, PLCC, RMSE and Krasula performance evaluation, it can be observed from the table that the proposed PAD-Net achieve the best performance in terms of SROCC, PLCC, RMSE, AUC-DS, AUC-BW and CC. The results of Krasula performance criteria demonstrate that PAD-Net is the most promising metric in distinguishing stereo images with different qualities.

\begin{table}[ht]
	\begin{center}
		\captionsetup{justification=centering}
		\caption{\textsc{Results of the T-Test on the LIVE Phase I and II Database.}}
		\label{table11}
		\scalebox{0.65}{
			\begin{tabular}{@{}c|cccc@{}}
				\toprule
				LIVE Phase I / II& Cyclopean MS-SSIM \cite{chen2013full}& CNN \cite{kang2014convolutional}& StereoQA-Net \cite{zhou2019dual}& Proposed PAD-Net \\ \midrule
				Cyclopean MS-SSIM \cite{chen2013full}& 0 / 0  & -1 / 1 & -1 / -1    & -1 / -1           \\
				CNN \cite{kang2014convolutional}& 1 / -1  & 0 / 0   & -1 / -1   & -1 / -1         \\
				StereoQA-Net \cite{zhou2019dual}& 1 / 1  & 1 / 1   & 0 / 0     & -1 / -1              \\
				Proposed PAD-Net  & 1 / 1     & 1 / 1   & 1 / 1   & 0 / 0          \\ \bottomrule
		\end{tabular}}
	\end{center}
\end{table}

Moreover, we conduct significance t-tests using the PLCC values of 10 runs \cite{oh2017blind} to verify whether our proposed model is statistically better than other metrics. Table \ref{table11} list the results of  t-tests on LIVE Phase I and II where ‘1’ or ‘-1’ indicate that the metric in the row is statistically superior or worse than the competitive metric in the column. The number ‘0’ means that the two metrics are statistically indistinguishable. From Table \ref{table11}, we can see that our proposed metric is statistically better than other metrics both on LIVE Phase I and II.

\subsection{Performance Evaluation on Individual Distortion Type}

\begin{figure}[ht]
	\centering
	\subfigure[]{
		\includegraphics[width=4.1cm]{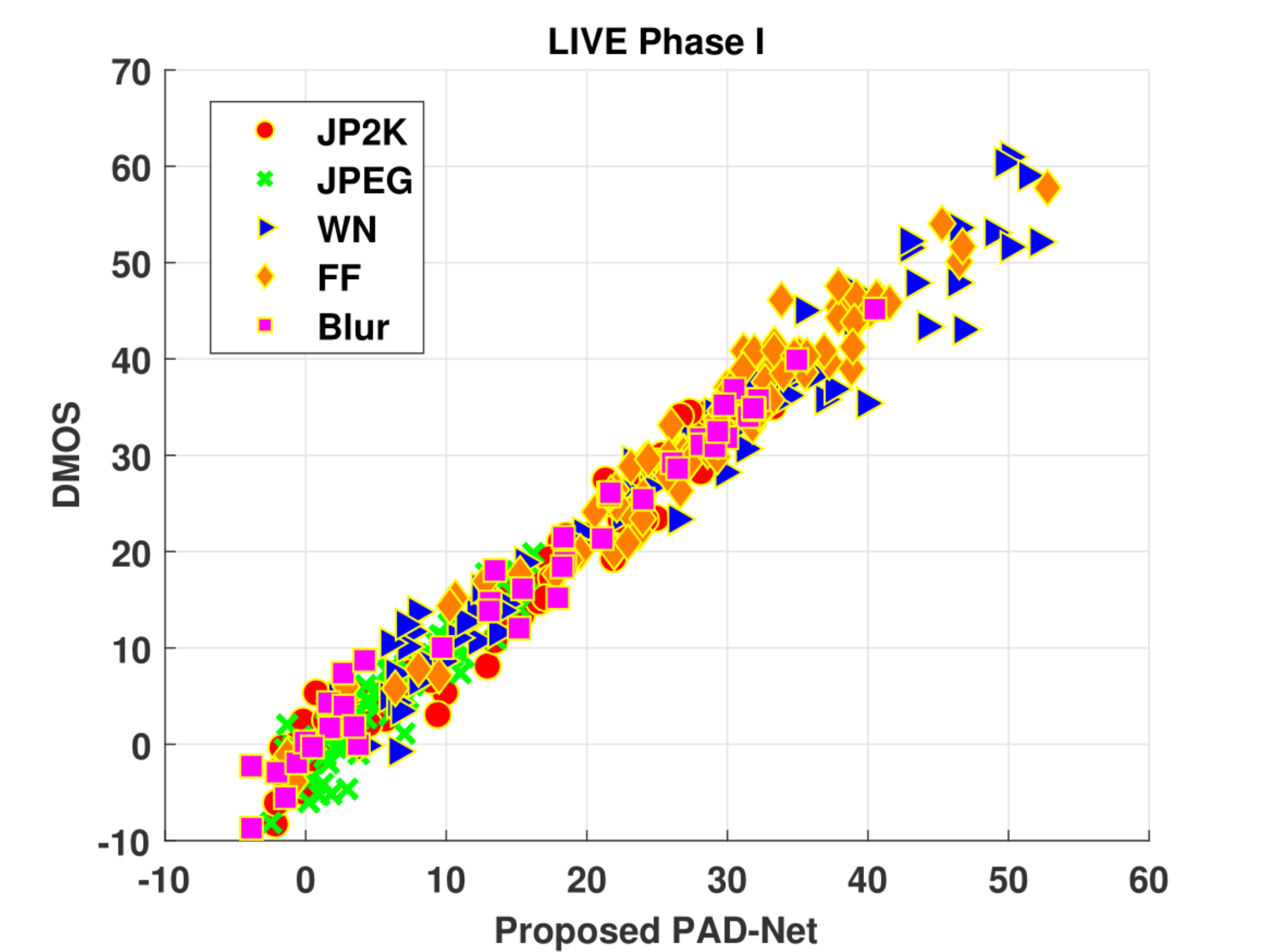}
	}
	\subfigure[]{
		\includegraphics[width=4.1cm]{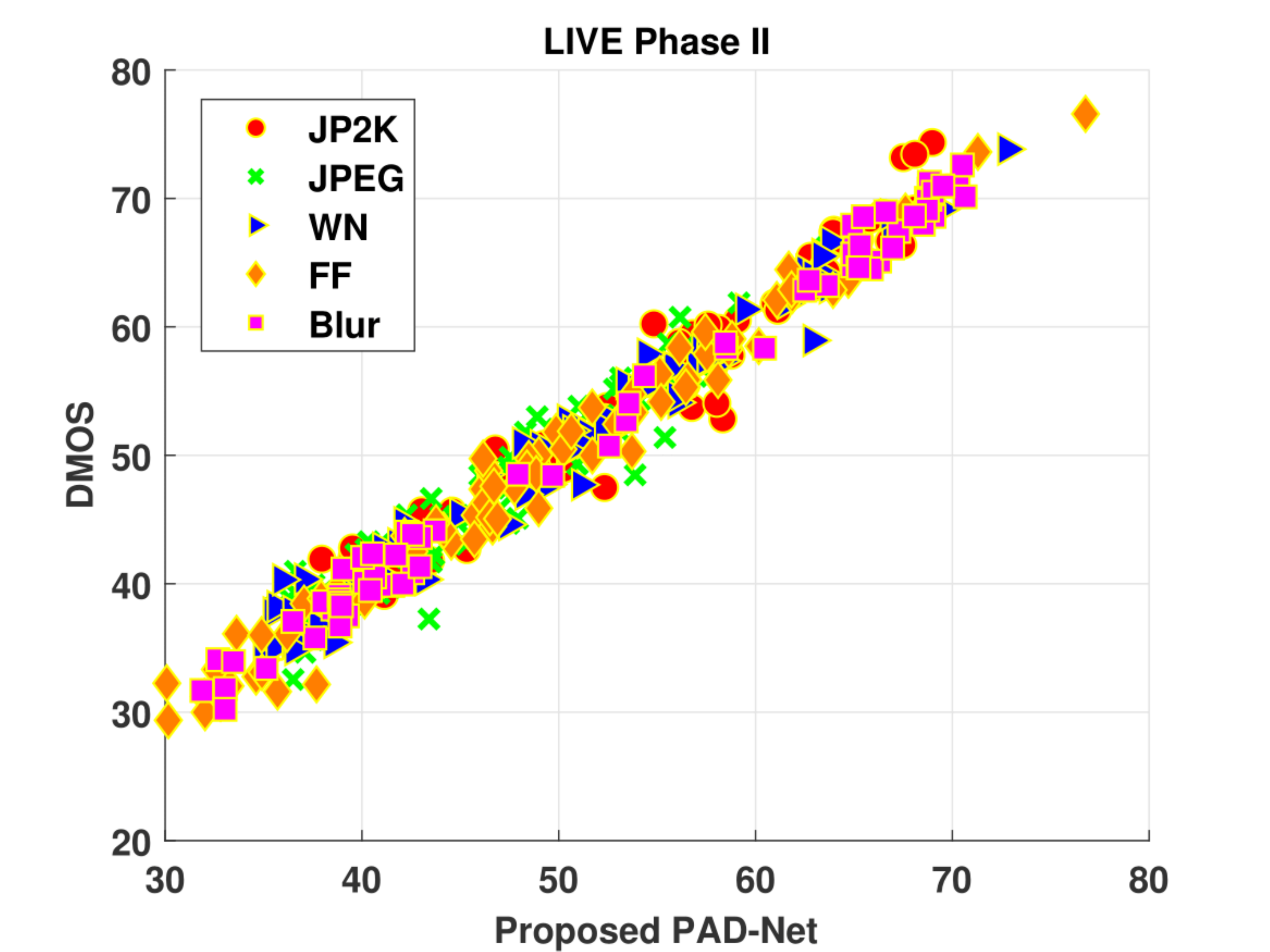}
	}
	\caption{Scatter plots of DMOS values against predictions by PAD-Net for individual distortion type on (a) LIVE Phase I and (b) LIVE Phase II.}
	\label{fig:fig7}
\end{figure}

We further investigate the capacity of our proposed PAD-Net for each distortion type, the SROCC and PLCC performance are illustrated in Table \ref{table6} and \ref{table7}. The best performing results across listed metrics are highlighted in boldface. As shown in Table \ref{table6} and \ref{table7}, our proposed model achieves competitive performance for most of the distortion types. In addition, the scatter plots of DMOS values versus objective scores predicted by PAD-Net for each distortion type on LIVE Phase I and II are presented in Fig. \ref{fig:fig7}(a) and \ref{fig:fig7}(b). The linear correlation between DMOS values and predicted scores demonstrates the great monotonicity and accuracy of PAD-Net. DMOS value range of JPEG compressed images is roughly narrower than those of other distortion types, making it more difficult to estimate the perceptual image quality. Thus, the PLCC and SROCC performance for JPEG distortion is generally lower than the other four.

\begin{table*}[htbp]
	\begin{center}
		\captionsetup{justification=centering}
		\caption{\textsc{SROCC Performance Comparison for Individual Distortion Type on LIVE I and II Databases.}}
		\label{table6}
		\setlength{\tabcolsep}{4.5mm}{
			\scalebox{0.88}{
				\begin{tabular}{@{}c|c|ccccc|ccccc@{}}
					\toprule
					&  & \multicolumn{5}{c|}{LIVE Phase I} & \multicolumn{5}{c}{LIVE Phase II} \\ \midrule
					Type & Metrics & JP2K & JPEG & WN & BLUR & FF & JP2K & JPEG & WN & BLUR & FF \\ \midrule
					\multirow{5}{*}{FR} & Gorley \cite{gorley2008stereoscopic} & 0.015 & 0.569 & 0.741 & 0.750 & 0.366 & 0.110 & 0.027 & 0.875 & 0.770 & 0.601 \\
					& You \cite{you2010perceptual} & 0.860 & 0.439 & 0.940 & 0.882 & 0.588 & 0.894 & 0.795 & 0.909 & 0.813 & 0.891 \\
					& Benoit \cite{benoit2009quality} & 0.910 & 0.603 & 0.930 & 0.931 & 0.699 & 0.751 & 0.867 & 0.923 & 0.455 & 0.773 \\
					& Lin \cite{lin2014quality} & 0.839 & 0.207 & 0.928 & \textbf{0.935} & 0.658 & 0.718 & 0.613 & 0.907 & 0.711 & 0.701 \\
					& Cyclopean MS-SSIM \cite{chen2013full} & 0.888 & 0.530 & 0.948 & 0.925 & 0.707 & 0.814 & 0.843 & 0.940 & 0.908 & 0.884 \\ \midrule
					\multirow{3}{*}{RR} & RR-BPI \cite{qi2015reduced} & - & - & - & - & - & 0.776 & 0.736 & 0.904 & 0.871 & 0.854 \\
					& RR-RDCT \cite{ma2016reorganized} & 0.887 & 0.616 & 0.912 & 0.879 & 0.696 & 0.879 & 0.737 & 0.732 & 0.876 & 0.895 \\
					& Ma \cite{ma2017reduced} & 0.907 & 0.660 & 0.928 & 0.921 & 0.792 & 0.868 & 0.791 & 0.954 & \textbf{0.923} & 0.944 \\ \midrule
					\multirow{9}{*}{NR} & Akhter \cite{akhter2010no} & 0.866 & 0.675 & 0.914 & 0.555 & 0.640 & 0.724 & 0.649 & 0.714 & 0.682 & 0.559 \\
					& Sazzad \cite{sazzad2012objective} & 0.721 & 0.526 & 0.807 & 0.597 & 0.705 & 0.625 & 0.479 & 0.647 & 0.775 & 0.725 \\
					& Chen \cite{chen2013no} & 0.863 & 0.617 & 0.919 & 0.878 & 0.652 & 0.867 & 0.867 & 0.950 & 0.900 & 0.933 \\
					& S3D-BLINQ \cite{su2015oriented} & - & - & - & - & - & 0.845 & 0.818 & 0.946 & 0.903 & 0.899 \\
					& CNN \cite{kang2014convolutional} & 0.857 & 0.477 & 0.874 & 0.782 & 0.670 & 0.660 & 0.598 & 0.769 & 0.317 & 0.476 \\
					& DNR-S3DIQE \cite{oh2017blind} & 0.885 & 0.765 & 0.921 & 0.930 & 0.944 & 0.853 & 0.822 & 0.833 & 0.889 & 0.878 \\
					& DBN \cite{yang2019blind} & 0.897 & 0.768 & 0.929 & 0.917 & 0.685 & 0.859 & 0.806 & 0.864 & 0.834 & 0.877 \\
					& StereoQA-Net \cite{zhou2019dual} & 0.961 & \textbf{0.912} & 0.965 & 0.855 & 0.917 & 0.874 & 0.747 & 0.942 & 0.600 & \textbf{0.951} \\
					& Proposed PAD-Net & \textbf{0.969} & 0.889 & \textbf{0.968} & 0.917 & \textbf{0.996} & \textbf{0.959} & \textbf{0.882} & \textbf{0.962} & 0.867 & 0.945 \\ \bottomrule
		\end{tabular}}}
	\end{center}
\end{table*}

\begin{table*}[htbp]
	\begin{center}
		\captionsetup{justification=centering}
		\caption{\textsc{PLCC Performance Comparison for Individual Distortion Type on LIVE I and II Databases.}}
		\label{table7}
		\setlength{\tabcolsep}{4.5mm}{
			\scalebox{0.88}{
				\begin{tabular}{@{}c|c|ccccc|ccccc@{}}
					\toprule
					&  & \multicolumn{5}{c|}{LIVE Phase I} & \multicolumn{5}{c}{LIVE Phase II} \\ \midrule
					Type & Metrics & JP2K & JPEG & WN & BLUR & FF & JP2K & JPEG & WN & BLUR & FF \\ \midrule
					\multirow{5}{*}{FR} & Gorley \cite{gorley2008stereoscopic} & 0.485 & 0.312 & 0.796 & 0.852 & 0.364 & 0.372 & 0.322 & 0.874 & 0.934 & 0.706 \\
					& You \cite{you2010perceptual} & 0.877 & 0.487 & 0.941 & 0.919 & 0.730 & 0.905 & 0.830 & 0.912 & 0.784 & 0.915 \\
					& Benoit \cite{benoit2009quality} & 0.939 & 0.640 & 0.925 & 0.948 & 0.747 & 0.784 & 0.853 & 0.926 & 0.535 & 0.807 \\
					& Lin \cite{lin2014quality} & 0.799 & 0.196 & 0.925 & 0.811 & 0.700 & 0.744 & 0.583 & 0.909 & 0.671 & 0.699 \\
					& Cyclopean MS-SSIM \cite{chen2013full} & 0.912 & 0.603 & 0.942 & 0.942 & 0.776 & 0.834 & 0.862 & 0.957 & 0.963 & 0.901 \\ \midrule
					\multirow{3}{*}{RR} & RR-BPI \cite{qi2015reduced} & - & - & - & - & - & 0.858 & 0.871 & 0.891 & 0.981 & 0.925 \\
					& RR-RDCT \cite{ma2016reorganized} & 0.918 & 0.722 & 0.913 & 0.925 & 0.807 & 0.897 & 0.748 & 0.810 & 0.969 & 0.910 \\
					& Ma \cite{ma2017reduced} & 0.940 & 0.720 & 0.935 & 0.936 & 0.843 & 0.880 & 0.765 & 0.932 & 0.913 & 0.906 \\ \midrule
					\multirow{9}{*}{NR} & Akhter \cite{akhter2010no} & 0.905 & 0.729 & 0.904 & 0.617 & 0.503 & 0.776 & 0.786 & 0.722 & 0.795 & 0.674 \\
					& Sazzad \cite{sazzad2012objective} & 0.774 & 0.565 & 0.803 & 0.628 & 0.694 & 0.645 & 0.531 & 0.657 & 0.721 & 0.727 \\
					& Chen \cite{chen2013no} & 0.907 & 0.695 & 0.917 & 0.917 & 0.735 & 0.899 & 0.901 & 0.947 & 0.941 & 0.932 \\
					& S3D-BLINQ \cite{su2015oriented} & - & - & - & - & - & 0.847 & 0.888 & 0.953 & 0.968 & 0.944 \\
					& CNN \cite{kang2014convolutional} & 0.956 & 0.630 & 0.983 & 0.862 & 0.846 & 0.685 & 0.567 & 0.855 & 0.455 & 0.662 \\
					& DNR-S3DIQE \cite{oh2017blind} & 0.913 & 0.767 & 0.910 & 0.950 & 0.954 & 0.865 & 0.821 & 0.836 & 0.934 & 0.915 \\
					& DBN \cite{yang2019blind} & 0.942 & 0.824 & 0.954 & 0.963 & 0.789 & 0.886 & 0.867 & 0.887 & 0.988 & 0.916 \\
					& StereoQA-Net \cite{zhou2019dual} & \textbf{0.988} & 0.916 & \textbf{0.988} & 0.974 & 0.965 & 0.905 & \textbf{0.933} & 0.972 & 0.955 & \textbf{0.994} \\
					& Proposed PAD-Net & 0.982 & \textbf{0.919} & 0.978 & \textbf{0.985} & \textbf{0.994} & \textbf{0.981} & 0.898 & \textbf{0.973} & \textbf{0.997} & 0.986 \\ \bottomrule
		\end{tabular}}}
	\end{center}
\end{table*}

\subsection{Performance Evaluation for Symmetric/Asymmetric Distortion}

\begin{table}[htbp]
	\begin{center}
		\captionsetup{justification=centering}
		\caption{\textsc{SROCC Performance for Symmetrically and Asymmetrically Distorted Image Pairs on LIVE Phase II and Waterloo IVC Phase I Databases. The Best Performing Results are Highlighted in Bold.}}
		\label{table5}
		\scalebox{0.75}{
			\begin{tabular}{@{}c|c|cc|cc@{}}
				\toprule
				&  & \multicolumn{2}{c|}{LIVE Phase II} & \multicolumn{2}{c}{Waterloo IVC Phase I} \\ \midrule
				Type & Metrics & Symmetric & Asymmetric & Symmetric & Asymmetric \\ \midrule
				\multirow{5}{*}{FR} & Gorley \cite{gorley2008stereoscopic} & 0.383 & 0.056 & 0.566 & 0.475 \\
				& You \cite{you2010perceptual} & 0.914 & 0.701 & 0.752 & 0.571 \\
				& Benoit \cite{benoit2009quality} & 0.860 & 0.671 & 0.728 & 0.577 \\
				& Lin \cite{lin2014quality} & 0.605 & 0.668 & 0.688 & 0.592 \\
				& Cyclopean MS-SSIM \cite{chen2013full} & 0.923 & 0.842 & 0.924 & 0.643 \\ \midrule
				\multirow{6}{*}{NR} & Akhter \cite{akhter2010no} & 0.420 & 0.517 & - & - \\
				& Chen \cite{chen2013no} & 0.918 & 0.834 & 0.934 & 0.907 \\
				& CNN \cite{kang2014convolutional} & 0.590 & 0.633 & - & - \\
				& S3D-BLINQ \cite{su2015oriented} & 0.937 & 0.849 & - & - \\
				& StereoQA-Net \cite{zhou2019dual} & 0.979 & 0.927 & 0.957 & 0.940 \\
				& Proposed PAD-Net & \textbf{0.982} & \textbf{0.954} & \textbf{0.985} & \textbf{0.978} \\ \bottomrule
		\end{tabular}}
	\end{center}
\end{table}

Our proposed PAD-Net is based on the predictive coding theory and applies deep neural networks to model binocular rivalry mechanism for better prediction of the stereo image quality. Binocular rivalry seldom happens in symmetrical distortion but plays an important role in asymmetrically distorted image quality measurement. Table \ref{table5} presents the SROCC performance for symmetrically and asymmetrically distorted images in LIVE Phase II and Waterloo IVC Phase I databases. PAD-Net demonstrates the extraordinary ability to predict the perceived quality of asymmetrically distorted stereo pairs by well consideration of the visual mechanism in binocular rivalry.

\begin{figure}[htbp]
	\centerline{\includegraphics[width=8cm]{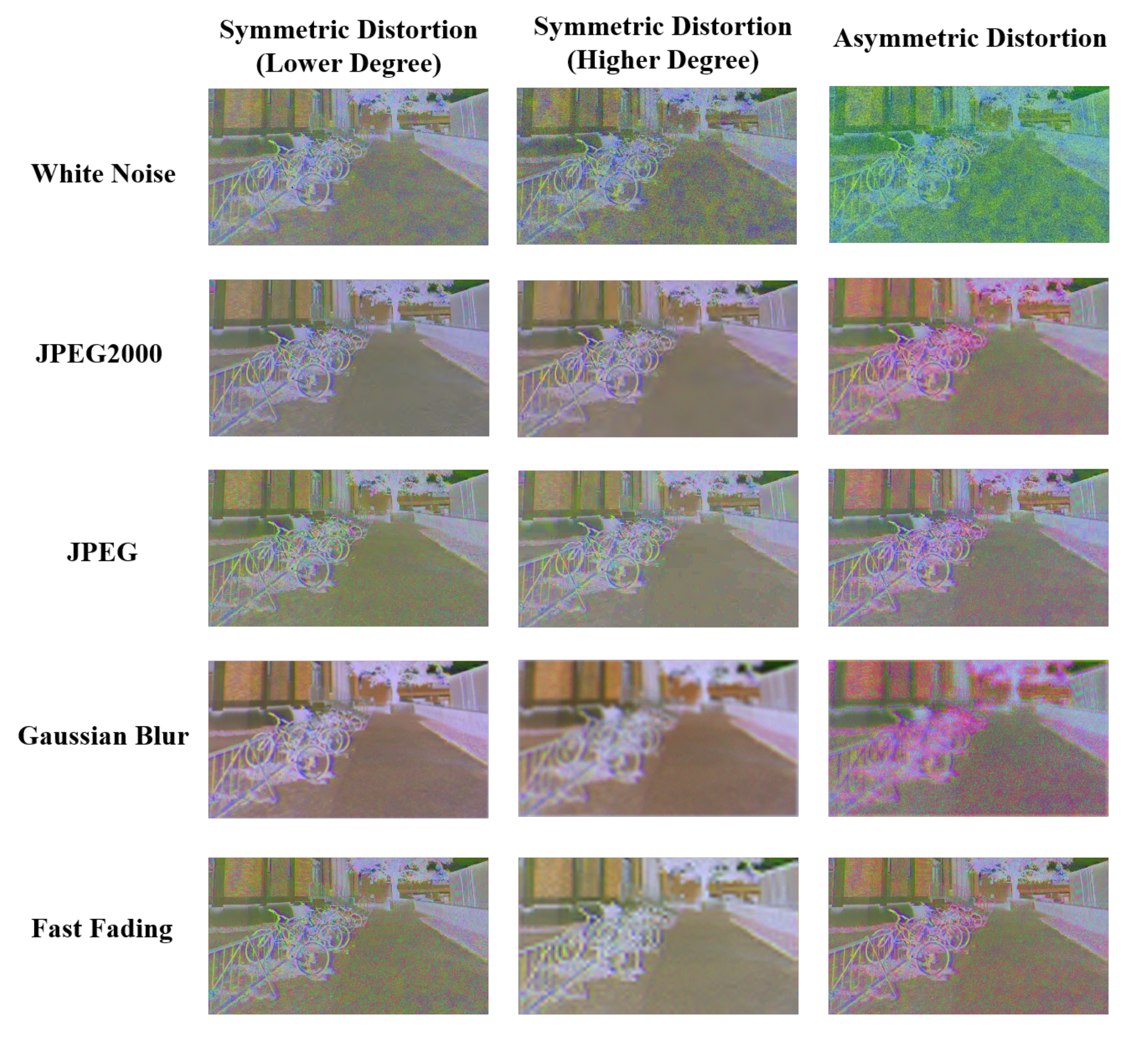}}
	\caption{Fusion maps for symmetrically and asymmetrically degraded stereo images under various distortion types.}
	\centering
	\label{fig:fig6}
\end{figure}

We provide some visualization results of the PAD-Net for better explanation. The fusion maps of distorted images, normalized prior and likelihood from left and right views are depicted in Fig. \ref{fig:fig6}. The colors of fusion map for symmetrically and asymmetrically distorted images are easy to be distinguished. For symmetrical distortions, the color of fusion maps is gray tone, which means the rivalry dominance for left and right views are similar. While for asymmetrical distortions, the fusion maps appear green or pink. 3D image quality is more affected by the poor-quality view for noise contamination as described in \cite{wang2015quality}. On the contrary, the perceptual quality is more affected by high-quality view for image blur \cite{wang2015quality}, thus the color tone of fusion maps for noise and blur are visually different. To be specific, for asymmetrically distorted images, white noise is different from the other four distortion types since noise tends to introduce high-frequency information while the other four distortion types are apt to remove details that correspond to high-frequency information.

\subsection{Cross Database Tests}

We conduct cross database tests to verify the generalization ability of our proposed PAD-Net. Models to be compared are trained on one database and tested on another. Table \ref{table8} presents the PLCC performance for cross database validation. Although PAD-Net does not show the best performance when trained on LIVE Phase II and tested on LIVE Phase I, it outperforms other metrics in the second round which is a more challenging task. Since LIVE Phase I only consists of symmetrical distortion while more than half of the 3D pairs in LIVE Phase II are asymmetrically distorted. PAD-Net trained on LIVE Phase I is able to handle the asymmetrical distortion in LIVE Phase II never met before. The PLCC performance on LIVE Phase II not only proves the generalization and robustness of PAD-Net but also demonstrates the effectiveness of the binocular rivalry mechanism based on predictive coding theory for asymmetric distortion in the proposed method.

\begin{table}[h]
	\begin{center}
		\captionsetup{justification=centering}
		\caption{\textsc{PLCC Performance of Cross Database Results. The Best Performing Results are Highlighted in Bold.}}
		\label{table8}
		\setlength{\tabcolsep}{2mm}{
		\scalebox{0.9}{
			\begin{tabular}{@{}c|c|c@{}}
				\toprule
				Metrics & Train LIVE II/Test LIVE I & Train LIVE I/Test LIVE II \\ \midrule
				Shao \cite{shao2016toward} & 0.894 & 0.779 \\
				Chen \cite{chen2013no} & 0.865 & - \\
				CNN \cite{kang2014convolutional} & 0.713 & 0.656 \\
				DBN \cite{yang2019blind} & 0.869 & 0.852 \\
				DECOSINE \cite{yang2018blind} & 0.916 & 0.846 \\
				StereoQA-Net \cite{zhou2019dual} & \textbf{0.932} & 0.710 \\
				Proposed PAD-Net & 0.915 & \textbf{0.854} \\ \bottomrule
		\end{tabular}}}
	\end{center}
\end{table}

\subsection{Effects of Network Structure}

\begin{table}[h]
	\begin{center}
		\captionsetup{justification=centering}
		\caption{\textsc{Performance Evaluation of Different Structure as Quality Regression Network. The Best Performing Results are Highlighted in Bold.}}
		\label{table9}
		\scalebox{0.85}{
			\begin{tabular}{@{}c|ccc|ccc@{}}
				\toprule
				& \multicolumn{3}{c|}{LIVE Phase I} & \multicolumn{3}{c}{LIVE Phase II} \\ \midrule
				Regression Structure & SROCC & PLCC & RMSE & SROCC & PLCC & RMSE \\ \midrule
				VGG-16 & 0.913 & 0.924 & 5.945 & 0.858 & 0.869 & 5.506 \\
				ResNet-18 & \textbf{0.973} & \textbf{0.975} & \textbf{3.514} & \textbf{0.967} & \textbf{0.975} & \textbf{2.446} \\
				ResNet-34 & 0.970 & 0.974 & 3.537 & 0.960 & 0.961 & 2.871 \\
				ResNet-50 & 0.968 & 0.974 & 3.588 & 0.963 & 0.965 & 2.727 \\ \bottomrule
		\end{tabular}}
	\end{center}
\end{table}

To explore the influence of different network structures as quality regression network, VGG-16 \cite{simonyan2014very}, ResNet-18, 34 and 50 \cite{he2016deep} are adopted to make comparisons. The SROCC, PLCC and RMSE performance on LIVE Phase I and II are reported in Table \ref{table9}. Firstly, ResNet has superior capability to extract discriminative features for quality prediction than VGG structure. Moreover, with the increased depth of ResNet, the performance does not improve. The possible explanation is that the limited training data requires shallow architecture. Generally, very deep networks need a huge amount of training data to achieve high performance. However, there are only hundreds of distorted images in LIVE Phase I and II, even with data augmentation, it is far from enough for deeper networks. Lack of training data may cause over-fitting problems for deeper neural networks. As a result, ResNet-18 is chosen in this paper to reach better tradeoff.

\subsection{Ablation Study}

Furthermore, ablation study is conducted to verify the effectiveness of each component in PAD-Net. We first feed the distorted left and right view images into quality regression module as the baseline. Then, normalized likelihood and prior maps are introduced to provide additive information for computing rivalry dominance of both views. Moreover, we compare different fusion methods.

\begin{figure}[htbp]
  \centerline{\includegraphics[width=9cm]{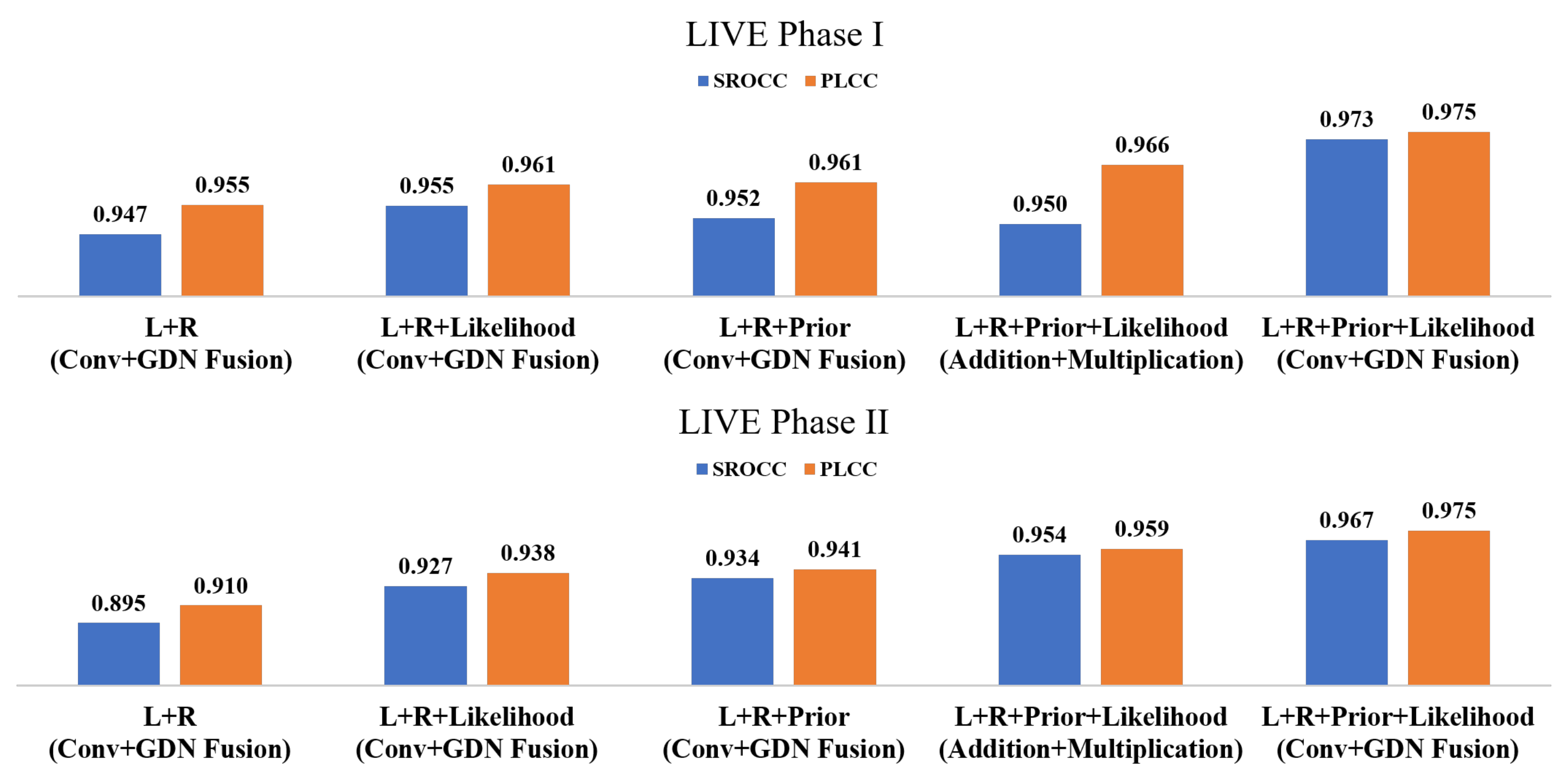}}
  \caption{Ablation Study on LIVE Phase I and Phase II Databases.}
  \centering
\label{fig:fig8}
\end{figure}

As shown in Fig. \ref{fig:fig8}, simply fusing left and right views can achieve promising performance on LIVE Phase I which only consists of symmetrically distorted pairs. However, the performance degrades seriously on LIVE Phase II owing to the existence of asymmetric distortion. According to the BRM-PC, prior and likelihood probability maps are necessary for 3D image quality estimation. The performance improvement on LIVE Phase II verify the effectiveness of prior and likelihood probability maps obtained through Siamese Encoder-decoder network and further demonstrate the superiority of the HVS guided binocular rivalry mechanism based on predictive coding theory. In addition, we compare the Conv+GDN fusion method with the intuitive addition+multiplication method which denotes we obtain the posterior probability by multiplying prior and likelihood probabilities. Note that Conv+GDN fusion means $GDN(Conv(Concat(\bm{P}_{nl},\bm{L}_{nl},\bm{I}_{l},\bm{P}_{nr},\bm{L}_{nr},\bm{I}_{r})))$ and Addition+Multiplication represents $\bm{P}_{nl}\times \bm{L}_{nl}\times \bm{I}_{l}+\bm{P}_{nr}\times \bm{L}_{nr}\times \bm{I}_{r}$, where $\bm{I}_{l}$ and $\bm{I}_{r}$ denote left and right view images, $\bm{P}_{nl}$, $\bm{P}_{nr}$, $\bm{L}_{nl}$, $\bm{L}_{nr}$ indicate the normalized prior and likelihood maps for both views. It is shown in Fig. \ref{fig:fig8} that our proposed method benefits a lot from the Conv+GDN fusion method since the parameters of fusion operation are updated during the training stage to generate the most discriminative feature maps for quality prediction. Therefore, the HVS guided Siamese encoder-decoder module to generate prior and likelihood map and the Conv+GDN fusion method are keys to the success of PAD-Net.

\subsection{Computation Complexity}
A good metric for blind SIQM should have high prediction accuracy as well as low computational cost. In the experiment, the models are tested on the NVIDIA GTX 1080ti GPU with 11GB memory. The running time for our proposed PAD-Net and other metrics are listed in Table \ref{table10}. Note that we record the time for predicting quality scores of 50 stereo images with the resolution of $360\times640$ and then average to obtain the time for each 3D image. The results in Table \ref{table10} show that PAD-Net only needs around 0.906 seconds per image which is significantly lower than other metrics.

\begin{table}[ht]
\begin{center}
\captionsetup{justification=centering}
\caption{\textsc{The Computation Time on NVIDIA GTX 1080ti GPU. The Best Performing Results are Highlighted in Bold.}}
\label{table10}
\scalebox{1.05}{
\begin{tabular}{@{}c|ccc@{}}
\toprule
Metrics & CNN \cite{kang2014convolutional}& StereoQA-Net \cite{zhou2019dual}& Proposed PAD-Net \\ \midrule
Time(sec) & 8.308 & 2.377 & \textbf{0.906} \\ \bottomrule
\end{tabular}}
\end{center}
\end{table}


\section{Conclusions}
In this paper, we explore a novel deep learning approach for blind stereoscopic image quality measurement according to the binocular rivalry mechanism based on predictive coding theory. Our proposed predictive auto-encoding network is an end-to-end architecture inspired by the human brain cognition process. Specifically, we adopt the Siamese encoder-decoder module to reconstruct binocular counterparts and generate the corresponding likelihood as well as prior maps. Moreover, we incorporate the quality regression module to obtain the final estimated perceptual quality score. The experimental results demonstrate that our proposed PAD-Net correlates well with subjective ratings. In addition, the proposed method outperforms state-of-the-art algorithms for distorted stereoscopic images under a variety of distortion types, especially for those with asymmetric distortions. Furthermore, we also show that the proposed PAD-Net has a promising generalization ability and can achieve lower time complexity. In future work, we intend to apply PredNet \cite{lotter2016deep} to mimic predictive coding theory in our framework and extend the method to blind stereoscopic video quality measurement. Except for image visual quality, we plan to investigate other 3D quality dimensions such as depth perception and visual comfort.

\bibliographystyle{IEEEtran}
\bibliography{references}

\begin{thebibliography}{10}
\providecommand{\url}[1]{#1}
\csname url@samestyle\endcsname
\providecommand{\newblock}{\relax}
\providecommand{\bibinfo}[2]{#2}
\providecommand{\BIBentrySTDinterwordspacing}{\spaceskip=0pt\relax}
\providecommand{\BIBentryALTinterwordstretchfactor}{4}
\providecommand{\BIBentryALTinterwordspacing}{\spaceskip=\fontdimen2\font plus
\BIBentryALTinterwordstretchfactor\fontdimen3\font minus
  \fontdimen4\font\relax}
\providecommand{\BIBforeignlanguage}[2]{{%
\expandafter\ifx\csname l@#1\endcsname\relax
\typeout{** WARNING: IEEEtran.bst: No hyphenation pattern has been}%
\typeout{** loaded for the language `#1'. Using the pattern for}%
\typeout{** the default language instead.}%
\else
\language=\csname l@#1\endcsname
\fi
#2}}
\providecommand{\BIBdecl}{\relax}
\BIBdecl

\bibitem{background}
\BIBentryALTinterwordspacing
``Stereoscopy.'' [Online]. Available:
  \url{https://en.wikipedia.org/wiki/Stereoscopy}
\BIBentrySTDinterwordspacing

\bibitem{xue2013matching}
T.~Xue, L.~Qu, and B.~Wu, ``Matching and 3-d reconstruction of multibubbles
  based on virtual stereo vision,'' \emph{IEEE Transactions on Instrumentation
  and Measurement}, vol.~63, no.~6, pp. 1639--1647, 2013.

\bibitem{anchini2006comparison}
R.~Anchini, C.~Liguori, V.~Paciello, and A.~Paolillo, ``A comparison between
  stereo-vision techniques for the reconstruction of 3-d coordinates of
  objects,'' \emph{IEEE Transactions on Instrumentation and Measurement},
  vol.~55, no.~5, pp. 1459--1466, 2006.

\bibitem{rajagopalan2004depth}
A.~Rajagopalan, S.~Chaudhuri, and U.~Mudenagudi, ``Depth estimation and image
  restoration using defocused stereo pairs,'' \emph{IEEE Transactions on
  Pattern Analysis and Machine Intelligence}, vol.~26, no.~11, pp. 1521--1525,
  2004.

\bibitem{allison2009binocular}
R.~S. Allison, B.~J. Gillam, and E.~Vecellio, ``Binocular depth discrimination
  and estimation beyond interaction space,'' \emph{Journal of Vision}, vol.~9,
  no.~1, pp. 10--10, 2009.

\bibitem{guo2014integrated}
Y.~Guo, M.~Bennamoun, F.~Sohel, M.~Lu, and J.~Wan, ``An integrated framework
  for 3-d modeling, object detection, and pose estimation from point-clouds,''
  \emph{IEEE Transactions on Instrumentation and Measurement}, vol.~64, no.~3,
  pp. 683--693, 2014.

\bibitem{de2008vector}
A.~De~Angelis, A.~Moschitta, F.~Russo, and P.~Carbone, ``A vector approach for
  image quality assessment and some metrological considerations,'' \emph{IEEE
  Transactions on Instrumentation and Measurement}, vol.~58, no.~1, pp. 14--25,
  2008.

\bibitem{angrisani2013internet}
L.~Angrisani, D.~Capriglione, L.~Ferrigno, and G.~Miele, ``An internet protocol
  packet delay variation estimator for reliable quality assessment of
  video-streaming services,'' \emph{IEEE Transactions on Instrumentation and
  Measurement}, vol.~62, no.~5, pp. 914--923, 2013.

\bibitem{yue2018effective}
G.~Yue, C.~Hou, T.~Zhou, and X.~Zhang, ``Effective and efficient blind quality
  evaluator for contrast distorted images,'' \emph{IEEE Transactions on
  Instrumentation and Measurement}, vol.~68, no.~8, pp. 2733--2741, 2018.

\bibitem{jiang2020blind}
Q.~Jiang, W.~Gao, S.~Wang, G.~Yue, F.~Shao, Y.-S. Ho, and S.~Kwong, ``Blind
  image quality measurement by exploiting high order statistics with deep
  dictionary encoding network,'' \emph{IEEE Transactions on Instrumentation and
  Measurement}, 2020.

\bibitem{jiang2020full}
Q.~Jiang, W.~Zhou, X.~Chai, G.~Yue, F.~Shao, and Z.~Chen, ``A full-reference
  stereoscopic image quality measurement via hierarchical deep feature
  degradation fusion,'' \emph{IEEE Transactions on Instrumentation and
  Measurement}, 2020.

\bibitem{russo2005automatic}
F.~Russo, ``Automatic enhancement of noisy images using objective evaluation of
  image quality,'' \emph{IEEE transactions on Instrumentation and Measurement},
  vol.~54, no.~4, pp. 1600--1606, 2005.

\bibitem{que2019exposure}
Y.~Que, Y.~Yang, and H.~J. Lee, ``Exposure measurement and fusion via adaptive
  multiscale edge-preserving smoothing,'' \emph{IEEE Transactions on
  Instrumentation and Measurement}, vol.~68, no.~12, pp. 4663--4674, 2019.

\bibitem{series2012subjective}
B.~Series, ``Subjective methods for the assessment of stereoscopic 3dtv
  systems,'' 2012.

\bibitem{chen2017blind}
Z.~Chen, W.~Zhou, and W.~Li, ``Blind stereoscopic video quality assessment:
  From depth perception to overall experience,'' \emph{IEEE Transactions on
  Image Processing}, vol.~27, no.~2, pp. 721--734, 2017.

\bibitem{shao2016toward}
F.~Shao, W.~Tian, W.~Lin, G.~Jiang, and Q.~Dai, ``Toward a blind deep quality
  evaluator for stereoscopic images based on monocular and binocular
  interactions,'' \emph{IEEE Transactions on Image Processing}, vol.~25, no.~5,
  pp. 2059--2074, 2016.

\bibitem{zhou20163d}
W.~Zhou, N.~Liao, Z.~Chen, and W.~Li, ``3d-hevc visual quality assessment:
  Database and bitstream model,'' in \emph{2016 Eighth International Conference
  on Quality of Multimedia Experience (QoMEX)}.\hskip 1em plus 0.5em minus
  0.4em\relax IEEE, 2016, pp. 1--6.

\bibitem{wang2016perceptual}
J.~Wang, S.~Wang, K.~Ma, and Z.~Wang, ``Perceptual depth quality in distorted
  stereoscopic images,'' \emph{IEEE Transactions on Image Processing}, vol.~26,
  no.~3, pp. 1202--1215, 2016.

\bibitem{chen2017visual}
J.~Chen, J.~Zhou, J.~Sun, and A.~C. Bovik, ``Visual discomfort prediction on
  stereoscopic 3d images without explicit disparities,'' \emph{Signal
  Processing: Image Communication}, vol.~51, pp. 50--60, 2017.

\bibitem{benoit2009quality}
A.~Benoit, P.~Le~Callet, P.~Campisi, and R.~Cousseau, ``Quality assessment of
  stereoscopic images,'' \emph{EURASIP journal on image and video processing},
  vol. 2008, no.~1, p. 659024, 2009.

\bibitem{you2010perceptual}
J.~You, L.~Xing, A.~Perkis, and X.~Wang, ``Perceptual quality assessment for
  stereoscopic images based on 2d image quality metrics and disparity
  analysis,'' in \emph{Proc. Int. Workshop Video Process. Quality Metrics
  Consum. Electron}, vol.~9, 2010, pp. 1--6.

\bibitem{gorley2008stereoscopic}
P.~Gorley and N.~Holliman, ``Stereoscopic image quality metrics and
  compression,'' in \emph{Stereoscopic Displays and Applications XIX}, vol.
  6803.\hskip 1em plus 0.5em minus 0.4em\relax International Society for Optics
  and Photonics, 2008, p. 680305.

\bibitem{chen2013full}
M.-J. Chen, C.-C. Su, D.-K. Kwon, L.~K. Cormack, and A.~C. Bovik,
  ``Full-reference quality assessment of stereopairs accounting for rivalry,''
  \emph{Signal Processing: Image Communication}, vol.~28, no.~9, pp.
  1143--1155, 2013.

\bibitem{lin2014quality}
Y.-H. Lin and J.-L. Wu, ``Quality assessment of stereoscopic 3d image
  compression by binocular integration behaviors,'' \emph{IEEE transactions on
  Image Processing}, vol.~23, no.~4, pp. 1527--1542, 2014.

\bibitem{qi2015reduced}
F.~Qi, D.~Zhao, and W.~Gao, ``Reduced reference stereoscopic image quality
  assessment based on binocular perceptual information,'' \emph{IEEE
  Transactions on multimedia}, vol.~17, no.~12, pp. 2338--2344, 2015.

\bibitem{ma2016reorganized}
L.~Ma, X.~Wang, Q.~Liu, and K.~N. Ngan, ``Reorganized dct-based image
  representation for reduced reference stereoscopic image quality assessment,''
  \emph{Neurocomputing}, vol. 215, pp. 21--31, 2016.

\bibitem{ma2017reduced}
J.~Ma, P.~An, L.~Shen, and K.~Li, ``Reduced-reference stereoscopic image
  quality assessment using natural scene statistics and structural
  degradation,'' \emph{IEEE Access}, vol.~6, pp. 2768--2780, 2017.

\bibitem{akhter2010no}
R.~Akhter, Z.~P. Sazzad, Y.~Horita, and J.~Baltes, ``No-reference stereoscopic
  image quality assessment,'' in \emph{Stereoscopic Displays and Applications
  XXI}, vol. 7524.\hskip 1em plus 0.5em minus 0.4em\relax International Society
  for Optics and Photonics, 2010, p. 75240T.

\bibitem{sazzad2012objective}
Z.~Sazzad, R.~Akhter, J.~Baltes, and Y.~Horita, ``Objective no-reference
  stereoscopic image quality prediction based on 2d image features and relative
  disparity,'' \emph{Advances in Multimedia}, vol. 2012, p.~8, 2012.

\bibitem{chen2013no}
M.-J. Chen, L.~K. Cormack, and A.~C. Bovik, ``No-reference quality assessment
  of natural stereopairs,'' \emph{IEEE Transactions on Image Processing},
  vol.~22, no.~9, pp. 3379--3391, 2013.

\bibitem{su2015oriented}
C.-C. Su, L.~K. Cormack, and A.~C. Bovik, ``Oriented correlation models of
  distorted natural images with application to natural stereopair quality
  evaluation,'' \emph{IEEE Transactions on image processing}, vol.~24, no.~5,
  pp. 1685--1699, 2015.

\bibitem{oh2017blind}
H.~Oh, S.~Ahn, J.~Kim, and S.~Lee, ``Blind deep s3d image quality evaluation
  via local to global feature aggregation,'' \emph{IEEE Transactions on Image
  Processing}, vol.~26, no.~10, pp. 4923--4936, 2017.

\bibitem{yang2019blind}
J.~Yang, Y.~Zhao, Y.~Zhu, H.~Xu, W.~Lu, and Q.~Meng, ``Blind assessment for
  stereo images considering binocular characteristics and deep perception map
  based on deep belief network,'' \emph{Information Sciences}, vol. 474, pp.
  1--17, 2019.

\bibitem{zhou2019dual}
W.~Zhou, Z.~Chen, and W.~Li, ``Dual-stream interactive networks for
  no-reference stereoscopic image quality assessment,'' \emph{IEEE Transactions
  on Image Processing}, 2019.

\bibitem{wang2004image}
Z.~Wang, A.~C. Bovik, H.~R. Sheikh, E.~P. Simoncelli \emph{et~al.}, ``Image
  quality assessment: from error visibility to structural similarity,''
  \emph{IEEE transactions on image processing}, vol.~13, no.~4, pp. 600--612,
  2004.

\bibitem{wang2002universal}
Z.~Wang and A.~C. Bovik, ``A universal image quality index,'' \emph{IEEE signal
  processing letters}, vol.~9, no.~3, pp. 81--84, 2002.

\bibitem{carnec2003image}
M.~Carnec, P.~Le~Callet, and D.~Barba, ``An image quality assessment method
  based on perception of structural information,'' in \emph{Proceedings 2003
  International Conference on Image Processing (Cat. No. 03CH37429)},
  vol.~3.\hskip 1em plus 0.5em minus 0.4em\relax IEEE, 2003, pp. III--185.

\bibitem{wang2005reduced}
Z.~Wang and E.~P. Simoncelli, ``Reduced-reference image quality assessment
  using a wavelet-domain natural image statistic model,'' in \emph{Human Vision
  and Electronic Imaging X}, vol. 5666.\hskip 1em plus 0.5em minus 0.4em\relax
  International Society for Optics and Photonics, 2005, pp. 149--159.

\bibitem{campisi2007stereoscopic}
P.~Campisi, P.~Le~Callet, and E.~Marini, ``Stereoscopic images quality
  assessment,'' in \emph{2007 15th European Signal Processing
  Conference}.\hskip 1em plus 0.5em minus 0.4em\relax IEEE, 2007, pp.
  2110--2114.

\bibitem{simonyan2014very}
K.~Simonyan and A.~Zisserman, ``Very deep convolutional networks for
  large-scale image recognition,'' \emph{arXiv preprint arXiv:1409.1556}, 2014.

\bibitem{he2016deep}
K.~He, X.~Zhang, S.~Ren, and J.~Sun, ``Deep residual learning for image
  recognition,'' in \emph{Proceedings of the IEEE conference on computer vision
  and pattern recognition}, 2016, pp. 770--778.

\bibitem{zhou2020blind}
W.~Zhou, Q.~Jiang, Y.~Wang, Z.~Chen, and W.~Li, ``Blind quality assessment for
  image superresolution using deep two-stream convolutional networks,''
  \emph{Information Sciences}, 2020.

\bibitem{yang2018blind}
J.~Yang, K.~Sim, X.~Gao, W.~Lu, Q.~Meng, and B.~Li, ``A blind stereoscopic
  image quality evaluator with segmented stacked autoencoders considering the
  whole visual perception route,'' \emph{IEEE Transactions on Image
  Processing}, vol.~28, no.~3, pp. 1314--1328, 2018.

\bibitem{howard1995binocular}
I.~P. Howard, B.~J. Rogers \emph{et~al.}, \emph{Binocular vision and
  stereopsis}.\hskip 1em plus 0.5em minus 0.4em\relax Oxford University Press,
  USA, 1995.

\bibitem{wang2015quality}
J.~Wang, A.~Rehman, K.~Zeng, S.~Wang, and Z.~Wang, ``Quality prediction of
  asymmetrically distorted stereoscopic 3d images,'' \emph{IEEE Transactions on
  Image Processing}, vol.~24, no.~11, pp. 3400--3414, 2015.

\bibitem{ohzawa1998mechanisms}
I.~Ohzawa, ``Mechanisms of stereoscopic vision: the disparity energy model,''
  \emph{Current opinion in neurobiology}, vol.~8, no.~4, pp. 509--515, 1998.

\bibitem{levelt1965binocular}
W.~J. Levelt, ``On binocular rivalry,'' Ph.D. dissertation, Van Gorcum Assen,
  1965.

\bibitem{dayan1998hierarchical}
P.~Dayan, ``A hierarchical model of binocular rivalry,'' \emph{Neural
  Computation}, vol.~10, no.~5, pp. 1119--1135, 1998.

\bibitem{hohwy2008predictive}
J.~Hohwy, A.~Roepstorff, and K.~Friston, ``Predictive coding explains binocular
  rivalry: An epistemological review,'' \emph{Cognition}, vol. 108, no.~3, pp.
  687--701, 2008.

\bibitem{spratling2017review}
M.~W. Spratling, ``A review of predictive coding algorithms,'' \emph{Brain and
  cognition}, vol. 112, pp. 92--97, 2017.

\bibitem{hume2003treatise}
D.~Hume, \emph{A treatise of human nature}.\hskip 1em plus 0.5em minus
  0.4em\relax Courier Corporation, 2003.

\bibitem{rao1999predictive}
R.~P. Rao and D.~H. Ballard, ``Predictive coding in the visual cortex: a
  functional interpretation of some extra-classical receptive-field effects,''
  \emph{Nature neuroscience}, vol.~2, no.~1, p.~79, 1999.

\bibitem{makhoul1975linear}
J.~Makhoul, ``Linear prediction: A tutorial review,'' \emph{Proceedings of the
  IEEE}, vol.~63, no.~4, pp. 561--580, 1975.

\bibitem{leopold1996activity}
D.~A. Leopold and N.~K. Logothetis, ``Activity changes in early visual cortex
  reflect monkeys' percepts during binocular rivalry,'' \emph{Nature}, vol.
  379, no. 6565, pp. 549--553, 1996.

\bibitem{chen2020stereoscopic}
Z.~Chen, J.~Xu, C.~Lin, and W.~Zhou, ``Stereoscopic omnidirectional image
  quality assessment based on predictive coding theory,'' \emph{IEEE Journal of
  Selected Topics in Signal Processing}, 2020.

\bibitem{friston2002functional}
K.~Friston, ``Functional integration and inference in the brain,''
  \emph{Progress in neurobiology}, vol.~68, no.~2, pp. 113--143, 2002.

\bibitem{kersten2004object}
D.~Kersten, P.~Mamassian, and A.~Yuille, ``Object perception as bayesian
  inference,'' \emph{Annu. Rev. Psychol.}, vol.~55, pp. 271--304, 2004.

\bibitem{ulyanov2018deep}
D.~Ulyanov, A.~Vedaldi, and V.~Lempitsky, ``Deep image prior,'' in
  \emph{Proceedings of the IEEE Conference on Computer Vision and Pattern
  Recognition}, 2018, pp. 9446--9454.

\bibitem{ma2017end}
K.~Ma, W.~Liu, K.~Zhang, Z.~Duanmu, Z.~Wang, and W.~Zuo, ``End-to-end blind
  image quality assessment using deep neural networks,'' \emph{IEEE
  Transactions on Image Processing}, vol.~27, no.~3, pp. 1202--1213, 2017.

\bibitem{ma2016waterloo}
K.~Ma, Z.~Duanmu, Q.~Wu, Z.~Wang, H.~Yong, H.~Li, and L.~Zhang, ``Waterloo
  exploration database: New challenges for image quality assessment models,''
  \emph{IEEE Transactions on Image Processing}, vol.~26, no.~2, pp. 1004--1016,
  2016.

\bibitem{sheikh2006statistical}
H.~R. Sheikh, M.~F. Sabir, and A.~C. Bovik, ``A statistical evaluation of
  recent full reference image quality assessment algorithms,'' \emph{IEEE
  Transactions on image processing}, vol.~15, no.~11, pp. 3440--3451, 2006.

\bibitem{moorthy2013subjective}
A.~K. Moorthy, C.-C. Su, A.~Mittal, and A.~C. Bovik, ``Subjective evaluation of
  stereoscopic image quality,'' \emph{Signal Processing: Image Communication},
  vol.~28, no.~8, pp. 870--883, 2013.

\bibitem{tong2006neural}
F.~Tong, M.~Meng, and R.~Blake, ``Neural bases of binocular rivalry,''
  \emph{Trends in cognitive sciences}, vol.~10, no.~11, pp. 502--511, 2006.

\bibitem{balle2016end}
J.~Ball{\'e}, V.~Laparra, and E.~P. Simoncelli, ``End-to-end optimized image
  compression,'' \emph{arXiv preprint arXiv:1611.01704}, 2016.

\bibitem{balle2018variational}
J.~Ball{\'e}, D.~Minnen, S.~Singh, S.~J. Hwang, and N.~Johnston, ``Variational
  image compression with a scale hyperprior,'' \emph{arXiv preprint
  arXiv:1802.01436}, 2018.

\bibitem{radford2015unsupervised}
A.~Radford, L.~Metz, and S.~Chintala, ``Unsupervised representation learning
  with deep convolutional generative adversarial networks,'' \emph{arXiv
  preprint arXiv:1511.06434}, 2015.

\bibitem{balle2015density}
J.~Ball{\'e}, V.~Laparra, and E.~P. Simoncelli, ``Density modeling of images
  using a generalized normalization transformation,'' \emph{arXiv preprint
  arXiv:1511.06281}, 2015.

\bibitem{summerfield2005mistaking}
C.~Summerfield, T.~Egner, J.~Mangels, and J.~Hirsch, ``Mistaking a house for a
  face: neural correlates of misperception in healthy humans,'' \emph{Cerebral
  Cortex}, vol.~16, no.~4, pp. 500--508, 2005.

\bibitem{glorot2011deep}
X.~Glorot, A.~Bordes, and Y.~Bengio, ``Deep sparse rectifier neural networks,''
  in \emph{Proceedings of the fourteenth international conference on artificial
  intelligence and statistics}, 2011, pp. 315--323.

\bibitem{nair2010rectified}
V.~Nair and G.~E. Hinton, ``Rectified linear units improve restricted boltzmann
  machines,'' in \emph{Proceedings of the 27th international conference on
  machine learning (ICML-10)}, 2010, pp. 807--814.

\bibitem{he2016identity}
K.~He, X.~Zhang, S.~Ren, and J.~Sun, ``Identity mappings in deep residual
  networks,'' in \emph{European conference on computer vision}.\hskip 1em plus
  0.5em minus 0.4em\relax Netherlands: Springer, 2016, pp. 630--645.

\bibitem{pan2009survey}
S.~J. Pan and Q.~Yang, ``A survey on transfer learning,'' \emph{IEEE
  Transactions on knowledge and data engineering}, vol.~22, no.~10, pp.
  1345--1359, 2009.

\bibitem{video2003final}
{Video Quality Experts Group} \emph{et~al.}, ``Final report from the video
  quality experts group on the validation of objective models of video quality
  assessment, phase ii,'' \emph{2003 VQEG}, 2003.

\bibitem{krasula2016accuracy}
L.~Krasula, K.~Fliegel, P.~Le~Callet, and M.~Kl{\'\i}ma, ``On the accuracy of
  objective image and video quality models: New methodology for performance
  evaluation,'' in \emph{2016 Eighth International Conference on Quality of
  Multimedia Experience (QoMEX)}.\hskip 1em plus 0.5em minus 0.4em\relax IEEE,
  2016, pp. 1--6.

\bibitem{mikolajczyk2018data}
A.~Miko{\l}ajczyk and M.~Grochowski, ``Data augmentation for improving deep
  learning in image classification problem,'' in \emph{2018 international
  interdisciplinary PhD workshop (IIPhDW)}.\hskip 1em plus 0.5em minus
  0.4em\relax IEEE, 2018, pp. 117--122.

\bibitem{kang2014convolutional}
L.~Kang, P.~Ye, Y.~Li, and D.~Doermann, ``Convolutional neural networks for
  no-reference image quality assessment,'' in \emph{Proceedings of the IEEE
  conference on computer vision and pattern recognition}, 2014, pp. 1733--1740.

\bibitem{lotter2016deep}
W.~Lotter, G.~Kreiman, and D.~Cox, ``Deep predictive coding networks for video
  prediction and unsupervised learning,'' \emph{arXiv preprint
  arXiv:1605.08104}, 2016.

\end{thebibliography}

%

%
%
%




\end{document}